%% file: main.tex
\pdfoutput=1
\documentclass[aps,prd,10pt,twocolumn,superscriptaddress,nofootinbib,floatfix,raggedbottom]{revtex4-2}
\usepackage{amsmath,amssymb,graphicx,booktabs,hyperref,xcolor}
\hypersetup{colorlinks=true,citecolor=blue,linkcolor=blue,urlcolor=blue,pdftitle={A high-precision Transformer model for precessing binary black-hole merger waveforms calibrated to numerical relativity},pdfauthor={Jameson Dong}}
\graphicspath{{figures/}}
\newcommand{\XPNR}{\textsc{IMRPhenomXPNR}}
\newcommand{\TPHM}{\textsc{IMRPhenomTPHM}}
\newcommand{\SEOB}{\textsc{SEOBNRv5PHM}}
\newcommand{\Sur}{\textsc{NRSur7dq4}}
\newcommand{\SurVTwo}{\textsc{NRSur7dq4v2}}
\input{tables/product_v6_macros}

\input{tables/product_v6_reference_macros}
\input{tables/v7_text_macros}

\input{tables/v7_runtime_macros}
\begin{document}
\title{A high-precision Transformer model for precessing binary black-hole merger waveforms calibrated to numerical relativity}
\author{Jameson Dong}\email{jdong34@jh.edu}\affiliation{William H. Miller III Department of Physics and Astronomy, Johns Hopkins University, Baltimore, MD 21218, USA}
\date{November 20, 2026}
\begin{abstract}
\input{sections/abstract}
\end{abstract}
\maketitle
\input{sections/introduction}
\input{sections/method}
\input{sections/data}
\input{sections/results}
\input{sections/limitations}
\input{sections/conclusion}
\input{sections/reproducibility}
\appendix
\input{sections/appendix_data}
\input{sections/appendix_seams}
\input{sections/appendix_method_checks}
\input{sections/appendix_reference_inputs}
\bibliography{bib/refs}
\end{document}

%% file: tables/product_v6_macros.tex
\providecommand{\ProductStrainMean}{0.0305}
\providecommand{\CommonProductStrainMean}{0.0301}
\providecommand{\ProductMismatchMean}{0.1536}
\providecommand{\CommonProductMismatchMean}{0.1496}
\providecommand{\ProductCurvatureMean}{0.1497}

\providecommand{\ProductRows}{109}
\providecommand{\ProductGroups}{106}

\providecommand{\CommonSEOBStrainMean}{0.1124}

\providecommand{\CommonSEOBMismatchMean}{1.5475}

\providecommand{\CommonNRSurStrainMean}{0.0679}

\providecommand{\CommonNRSurMismatchMean}{0.6049}

\providecommand{\CommonTPHMStrainMean}{0.1546}

\providecommand{\CommonTPHMMismatchMean}{2.5983}

%% file: tables/product_v6_reference_macros.tex
\providecommand{\ProductFixedLTwo}{0.2414}
\providecommand{\ProductPeakRotatedLTwo}{0.1338}
\providecommand{\ProductMedianAngleDeg}{3.16}
\providecommand{\ProductMedianPeakShift}{0.66}
\providecommand{\ProductFixedReferenceLTwo}{0.0939}

%% file: tables/v7_text_macros.tex
\providecommand{\VSevenCommonSixty}{0.0579}
\providecommand{\VSevenCommonOneEighty}{0.2358}
\providecommand{\SeobCommonSixty}{0.9315}
\providecommand{\SeobCommonOneEighty}{1.7449}
\providecommand{\SurCommonSixty}{0.4369}
\providecommand{\SurCommonOneEighty}{0.6949}
\providecommand{\TphmCommonSixty}{1.6388}
\providecommand{\TphmCommonOneEighty}{2.9404}
\providecommand{\VSevenInboxStrain}{0.0184}
\providecommand{\VSevenInboxCurvature}{0.1106}
\providecommand{\VSevenInboxMismatch}{0.0253}
\providecommand{\VSevenInboxHighMass}{0.0340}
\providecommand{\SurInboxStrain}{0.0240}
\providecommand{\SurInboxCurvature}{0.0856}
\providecommand{\SurInboxMismatch}{0.0326}
\providecommand{\SurInboxHighMass}{0.0338}
\providecommand{\VSevenInboxMedian}{0.0163}
\providecommand{\SurInboxMedian}{0.0082}
\providecommand{\VSevenInboxP}{0.0549}
\providecommand{\SurInboxP}{0.0831}
\providecommand{\VSevenRitAll}{0.0809}
\providecommand{\VSevenRitAligned}{0.0525}
\providecommand{\VSevenRitPrecessing}{0.1176}
\providecommand{\VSevenRitCommon}{0.0741}
\providecommand{\SurRitCommon}{0.0721}
\providecommand{\InboxStrainDiff}{-0.0056}
\providecommand{\InboxStrainDiffLo}{-0.0112}
\providecommand{\InboxStrainDiffHi}{-0.0011}
\providecommand{\EarlyExtStrain}{0.0252}
\providecommand{\LateExtStrain}{0.2420}
\providecommand{\FullDenseStrain}{0.0265}

\providecommand{\ExTwoFourTwoNineLW}{0.027}
\providecommand{\ExTwoFourTwoNineSur}{0.032}
\providecommand{\ExTwoFourTwoNineSEOB}{0.097}
\providecommand{\ExTwoFourTwoNineTPHM}{0.132}
\providecommand{\ExTwoSixTwoNineLW}{0.139}
\providecommand{\ExTwoSixTwoNineSEOB}{0.515}
\providecommand{\ExTwoSixTwoNineTPHM}{0.600}
\providecommand{\UnpaddedCommonSurMOneTwenty}{0.5817}
\providecommand{\UnpaddedCommonLWMOneTwenty}{0.1490}
\providecommand{\UnpaddedInboxSurMOneTwenty}{0.0333}
\providecommand{\UnpaddedInboxLWMOneTwenty}{0.0251}
\providecommand{\CoreRefSXSStrain}{0.0291}
\providecommand{\CoreRefSXSMismatch}{0.1470}
\providecommand{\CoreRefCommonStrain}{0.0287}

%% file: tables/v7_runtime_macros.tex
\providecommand{\CompiledMaxRel}{3.8\times 10^{-15}}

\providecommand{\CompiledGPUS}{2.301}

\providecommand{\CompiledGPUAmortizedMS}{2.25}
\providecommand{\CompiledGPUThroughput}{445}

\providecommand{\CompiledGPUMemory}{9.68}

\providecommand{\CompiledCPUBatch}{32}
\providecommand{\CompiledCPUThreads}{4}
\providecommand{\CompiledCPUBatchS}{7.75}

\providecommand{\CompiledCPUThroughput}{4.13}
\providecommand{\CompiledCPUAmortizedMS}{242.3}

%% file: sections/abstract.tex
We present a Transformer model for merger-centered gravitational waveforms from precessing binary black holes. Given the mass ratio and two three-dimensional spin vectors, the network predicts 512 quintic B-spline coefficients for all 21 signed spin-weighted spherical-harmonic strain modes through $\ell=4$ on $[-1620,160]M$ ($M$ being the total binary mass), enforcing binary-exchange and equatorial-reflection equivariance and yielding $C^4$ time-domain waveforms with exact parameter and temporal derivatives via automatic differentiation. The model is pretrained jointly on numerical-relativity (NR) waveforms from the SXS and RIT catalogues together with synthetic \Sur{}, \SEOB{}, and perturbative waveforms, and fine-tuned on SXS strain data. On \ProductRows\ SXS evaluation waveforms (\ProductGroups\ physical systems) over $[-1500,100]M$, it achieves a mean balanced relative strain error of \ProductStrainMean\ and a mean finite-window detector mismatch at $120\,M_\odot$ of \ProductMismatchMean\%, outperforming \SEOB{}, \SurVTwo{}, and \TPHM{} across common support, with compiled batched \texttt{float64} inference reaching $\CompiledGPUThroughput\text{ waveforms/s}$ on an A100 GPU.

%% file: sections/introduction.tex
\section{Introduction}\label{sec:intro}

Modeling the merger of a precessing black-hole binary requires predicting a hierarchy of spin-weighted spherical-harmonic modes. Precession of the orbital plane tilts the radiation axis, breaks the equatorial reflection symmetry between positive- and negative-$m$ modes~\cite{Boyle2014ModeAsymmetry}, and couples the orbital and precessional phases to all six spin components. Numerical relativity (NR) resolves these strong-field dynamics directly, with public waveform catalogues produced by the Simulating eXtreme Spacetimes (SXS)~\cite{Boyle2019SXS,Scheel2025SXS} and Rochester Institute of Technology (RIT)~\cite{Healy2022RIT} collaborations, with the BAM code~\cite{Bruegmann2008BAM,Hamilton2024BAMCatalog}, and in the MAYA catalogue~\cite{Ferguson2025MAYA}, but computational cost limits the density with which the seven-dimensional intrinsic parameter space can be sampled. Reduced-order surrogate models interpolate NR waveforms accurately inside bounded parameter domains~\cite{Field2014surrogate}, including the precessing surrogates \textsc{NRSur7dq2}~\cite{Blackman2017NRSur7dq2}, \Sur{}~\cite{Varma2019NRSur7dq4} (covering mass ratios $q=m_1/m_2\leq4$ and dimensionless spin magnitudes $|\boldsymbol\chi_{1,2}|\leq0.8$), and its multi-domain extension \SurVTwo{}~\cite{Ravishankar2026NRSur7dq4v2}. To cover the broader parameter space and arbitrarily low starting frequencies, semi-analytical approximants combine post-Newtonian, test-mass, and ringdown physics with NR calibration. These include effective-one-body (EOB) models such as \SEOB{}~\cite{RamosBuades2023SEOBNRv5PHM,Estelles2025Asym} and \textsc{TEOBResumS}~\cite{Gamba2022PrecessingEOB}, frequency-domain phenomenological models such as \textsc{IMRPhenomXPHM} and its SpinTaylor extension~\cite{Pratten2021XPHM,Colleoni2025SpinTaylor}, \textsc{PhenomPNR}~\cite{Hamilton2021PNR}, \XPNR{}~\cite{Hamilton2025XPNR}, and \textsc{IMRPhenomXO4a}~\cite{Thompson2024XO4a}, and native time-domain phenomenological models such as \TPHM{}~\cite{Estelles2022TPHM}.

Machine-learning and differentiable waveform representations offer a complementary route to fast waveform evaluation and exact automatic parameter differentiation~\cite{Chua2019ROMAN,Khan2021ANNsurrogate,Schmidt2021mlgw}, supporting both gradient-based Bayesian samplers such as \texttt{Jim}~\cite{Wong2023Jim} and simulation-based neural posterior estimation such as \textsc{Dingo}~\cite{Dax2021Dingo}. Differentiable implementations of analytical and reduced-order models include the frequency-domain \texttt{ripple} library~\cite{Edwards2024Ripple} and JAX-accelerated versions of the nonprecessing \texttt{mlgw} and \texttt{mlgw-bns} surrogates~\cite{FrattaleMascioli2026Acceleration}. Neural surrogate models have been constructed for aligned-spin higher modes~\cite{Grimbergen2024HigherModes}, restricted precessing binaries~\cite{Thomas2022PrecessingNN}, and merger-remnant properties~\cite{Thomas2025RemnantNN}. Agentic symbolic regression has been used to construct analytic surrogates of eccentric dominant-mode waveforms~\cite{Islam2026AgenticSurrogates}. Recent precessing extensions emulate \SEOB{} in a decomposed orbital-phase, rotating-frame-mode, and quaternion representation~\cite{Whittall2026NNSur7dq10}, predict the empirical-interpolation components of \Sur{}~\cite{Purrer2026NRSurNN}, or learn compressed inertial modes through $\ell=4$ from \Sur{}~\cite{Modrekiladze2026InertialNN}. Transformer architectures have also been applied to forecast aligned-spin merger-ringdown polarizations from preceding inspiral time series~\cite{Tiki2026TransformerModes}.

Pretraining on synthetic approximants followed by NR fine-tuning has been demonstrated for aligned-spin dominant and higher modes in the DANSur framework~\cite{Freitas2025DANSur,Freitas2026DANSurHM}. Here we extend this approach to seven-parameter precessing merger waveforms over $[-1620,160]M$, evaluated primarily on $[-1500,100]M$. A Transformer decoder maps the binary parameters to 512 quintic B-spline coefficients (with a two-sided analytic bump knot density) for all 21 signed inertial strain modes through $\ell=4$, using cross-attention to condition spline tokens on the binary inputs and self-attention to couple temporal locations and mode channels. Predicting the signed inertial modes directly in the reference frame defined by the binary's horizon geometry avoids decomposing the waveform into a coprecessing frame and a separate orbital-phase branch. A four-element group projection enforces exact binary-exchange and equatorial-reflection equivariance.

To cover the seven-dimensional parameter space before NR fine-tuning, we pretrain the network jointly on SXS and RIT NR waveforms together with synthetic \Sur{}, \SEOB{}, and perturbative waveforms, supervising all 21 modes where available and projecting onto the modeled mode subspace for generators providing only a subset of modes. A soft unaligned SXS anchor and fixed-azimuth \Sur{} supervision constrain the output time and azimuth relative to the input triad, while finite-difference penalties regularize early, merger, and late-ringdown derivatives. Subsequent SXS-only fine-tuning yields the final model, whose accuracy and compiled inference speed we evaluate against held-out NR waveforms, reference-frame diagnostics, and analytic Weyl curvature $\Psi_4 = -\partial_\tau^2 h$.

Section~\ref{sec:method} defines the waveform representation, symmetry projection, and training objectives. Section~\ref{sec:data} describes the NR datasets, synthetic supervision, and evaluation metrics. Section~\ref{sec:results} presents the strain, reference-frame, and curvature evaluations, and Sec.~\ref{sec:limitations} discusses parameter-space continuity and physical scope.

%% file: sections/method.tex
\section{Waveform model}\label{sec:method}

\subsection{Physical inputs and time interval}

We model dimensionless complex gravitational-wave strain $h = h_+ - i h_\times$, with plus and cross polarizations $h_+$ and $h_\times$, in an inertial frame as a spin-weighted spherical-harmonic expansion,
\begin{equation}
 h(\tau,\iota,\varphi;x)=\sum_{\ell=2}^{4}\sum_{m=-\ell}^{\ell}
 h_{\ell m}(\tau;x)\,{}_{-2}Y_{\ell m}(\iota,\varphi),
 \label{eq:mode_expansion}
\end{equation}
where ${}_{-2}Y_{\ell m}(\iota,\varphi)$ are the spin-weight $-2$ spherical harmonics along line-of-sight polar and azimuthal angles $(\iota,\varphi)$, $\tau=t/M$ is time in units of total binary mass $M=m_1+m_2$ for component masses $m_{1,2}$ ($G=c=1$), and
\begin{equation}
 x=(q,\boldsymbol\chi_1,\boldsymbol\chi_2)
\end{equation}
collects the seven intrinsic parameters: the mass ratio $q=m_1/m_2>0$ (conventionally quoted with $q\geq1$, smoothly extended to $q>0$ under body exchange, with symmetric mass ratio $\nu=m_1m_2/M^2=q/(1+q)^2$) and the three-dimensional dimensionless spin vectors $\boldsymbol\chi_{1,2}$ of the two black holes. All 21 signed modes through $\ell=4$, including odd-$m$ and $m=0$ modes, are predicted simultaneously from $x$; distance scaling and detector projection are applied subsequently.

The spin components are specified at the dimensionless reference frequency $Mf_{\rm ref}=(20\,\mathrm{Hz})(120\,M_\odot)G/c^3\simeq0.0118$, defined for the NR inputs from the reference orbital frequency $\Omega$ by $f_{\rm ref}=\Omega/\pi$ (Sec.~\ref{sec:data}). This nominal quadrupole frequency equals $20\text{ Hz}$ at $M=120\,M_\odot$ and scales as $f_{\rm ref}\propto M^{-1}$ with total mass. At that reference epoch, each spin vector $\boldsymbol\chi_a=(\chi_{ax},\chi_{ay},\chi_{az})$ ($a=1,2$) is resolved in the orthonormal triad $(\boldsymbol n,\hat{\boldsymbol L}\times\boldsymbol n,\hat{\boldsymbol L})$, where $\boldsymbol n$ points from body 2 to body 1, $\hat{\boldsymbol L}$ is the unit orbital angular momentum direction, $\boldsymbol\chi_{a\perp}=(\chi_{ax},\chi_{ay})$ is the in-plane spin, and $\chi_{az}=\boldsymbol\chi_a\cdot\hat{\boldsymbol L}$ is the aligned spin component.

The spline representation spans $\tau\in[-1620,160]M$ and can be evaluated at arbitrary times within this interval. Primary evaluations use the $[-1500,100]M$ comparison window, with the surrounding margins accommodating time shifts during alignment; summary scores on the shorter $[-1000,40]M$ core interval are also noted for reference.

\subsection{Transformer and spline representation}\label{sec:decoder}

\begin{figure*}[tp]
 \centering
 \includegraphics[width=0.94\textwidth]{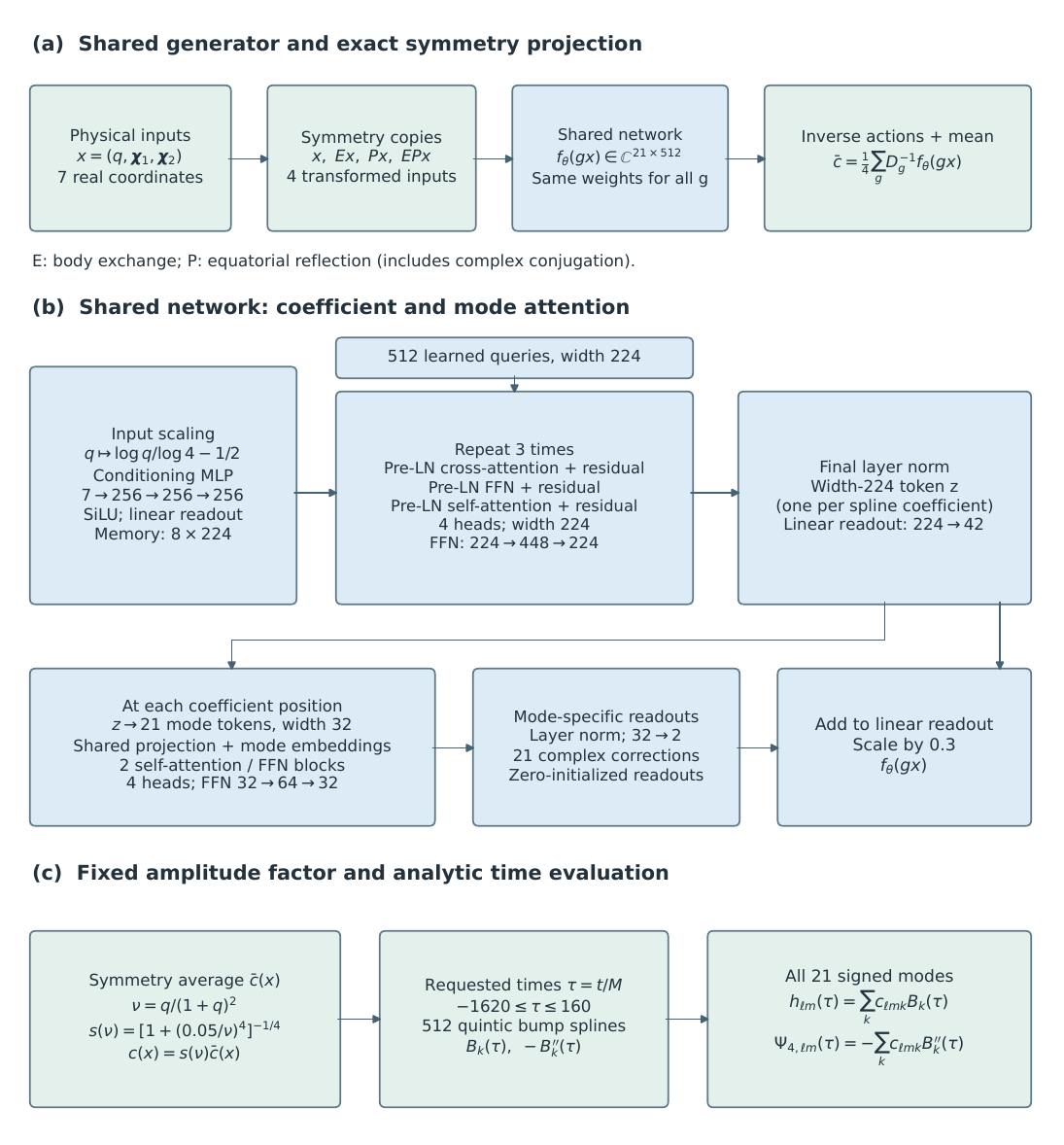}
 \caption{Parameter-to-waveform architecture. (a) Four symmetry-transformed copies of the binary input $x$ pass through a shared network $f_\theta$ and are averaged under the inverse group representation (Sec.~\ref{sec:gauge}). (b) Inside $f_\theta$, three Transformer blocks couple 512 spline-coefficient tokens to eight parameter-conditioning tokens and in time, followed by a two-block self-attention readout over the 21 mode channels at each spline index. (c) Multiplying by the symmetric-mass-ratio factor $s(\nu)$ (Sec.~\ref{sec:gauge}) yields the quintic B-spline coefficients $c_{\ell m k}(x)$, which determine both the strain modes $h_{\ell m}(\tau)$ on $[-1620,160]M$ and their analytic second time derivatives $\Psi_{4,\ell m}(\tau)=-\partial_\tau^2 h_{\ell m}(\tau)$.}
 \label{fig:architecture}
\end{figure*}

Figure~\ref{fig:architecture} illustrates the parameter-to-waveform architecture. A shared coefficient network $f_\theta$ with trainable parameters $\theta$ maps the scaled input features
\begin{equation}
 \left(\frac{\log q}{\log4}-\frac12,\boldsymbol\chi_1,\boldsymbol\chi_2\right)
 \label{eq:input_chart}
\end{equation}
through a three-layer multilayer perceptron (MLP) conditioner of width 256 with SiLU activations, followed by a linear projection to eight conditioning memory tokens of width 224. An attention decoder, adapted from the Transformer~\cite{Vaswani2017Attention}, operates on 512 learned query embeddings of width 224 through three sequential blocks. Each block applies pre-LayerNorm four-head cross-attention to the separately normalized conditioning tokens, a pre-LayerNorm feed-forward network (FFN) of hidden width 448 with GELU activation, and pre-LayerNorm four-head noncausal self-attention over all 512 coefficient tokens, each with a residual connection. Self-attention couples the spline tokens globally over the inspiral, merger, and ringdown intervals. After a final layer normalization, a shared linear readout maps each token to the 42 real and imaginary components of the 21 modes. A second, additive mode-attention branch couples the 21 modes at each spline index: it projects each normalized width-224 token to width 32, adds a learned mode embedding, and applies two pre-LayerNorm four-head self-attention blocks (with width-64 GELU feed-forward layers) over the 21 mode tokens. Mode-specific linear output layers, initialized to zero, map the normalized mode tokens to complex coefficient corrections, and the combined readout is scaled by a fixed factor of $0.3$ to temper initial output amplitudes before symmetry projection. The shared network has $2{,}564{,}276$ trainable parameters.

After applying the symmetry projection and mass-ratio factor defined in Sec.~\ref{sec:gauge}, the complex coefficients $c_{\ell m k}(x)$ define each inertial mode as a degree-five B-spline expansion~\cite{deBoor2001Splines},
\begin{equation}
 h_{\ell m}(\tau;x)=\sum_{k=1}^{512}c_{\ell m k}(x)B_k(\tau).
 \label{eq:model}
\end{equation}
The 512 basis functions $B_k(\tau)$ are determined by clamped endpoint knots of multiplicity six at $-1620M$ and $160M$ and 506 simple interior breakpoints spaced uniformly in the cumulative integral of the two-sided analytic bump knot-density profile
\begin{equation}
 \rho(\tau)=1+3\left[\sigma\!\left(\frac{\tau+200}{40}\right)-\sigma\!\left(\frac{\tau-75}{15}\right)\right],
 \label{eq:knot_density}
\end{equation}
with $\sigma(z)=(1+e^{-z})^{-1}$, concentrating four times higher knot resolution across the late inspiral, merger, and quasinormal-mode ringdown while smoothly relaxing back toward unit density across the $(100,160]M$ post-ringdown margin. Differentiating $B_k(\tau)$ twice with respect to $\tau$ yields the dimensionless Weyl curvature scalar mode expansion from the same coefficients,
\begin{equation}
 \Psi_{4,\ell m}(\tau;x)=-\partial_\tau^2 h_{\ell m}(\tau;x)=-\sum_{k=1}^{512}c_{\ell m k}(x)B_k''(\tau).
 \label{eq:psi4_model}
\end{equation}

For fixed network weights, the coefficient map $x\mapsto c_{\ell m k}(x)$ is real analytic on $q>0$. The strain and curvature have $C^4$ and $C^2$ continuity at interior time knots. These properties follow from the smooth network operations and the quintic basis; see Appendix~\ref{app:regularity}.

\subsection{Physical symmetries and alignment gauge}\label{sec:gauge}

In general relativity, discrete binary-exchange and parity transformations impose exact relations among the spherical-harmonic modes of symmetry-related binaries, providing a model-independent consistency diagnostic for precessing waveform generators~\cite{Estelles2026Parity}. In the separation-referenced frame $(\boldsymbol n,\hat{\boldsymbol L}\times\boldsymbol n,\hat{\boldsymbol L})$, exchanging the two bodies ($E$) reverses $\boldsymbol n$ and rotates the first two reference axes by $\pi$ about $\hat{\boldsymbol L}$. Equatorial reflection ($P$) reflects across the orbital plane ($\iota\mapsto\pi-\iota$) and flips the in-plane components of the axial spin vectors; the axial orbital normal $\hat{\boldsymbol L}$ is unchanged. Let $R=\operatorname{diag}(-1,-1,1)$. The two commuting involutions act on the input parameters $x$ as
\begin{align}
 Ex&=(q^{-1},R\boldsymbol\chi_2,R\boldsymbol\chi_1),\\
 Px&=(q,R\boldsymbol\chi_1,R\boldsymbol\chi_2),
 \label{eq:input_symmetry}
\end{align}
and on any 21-mode coefficient array $a=\{a_{\ell m k}\}$ as
\begin{align}
 (D_Ea)_{\ell m k}&=(-1)^m a_{\ell m k},\\
 (D_Pa)_{\ell m k}&=(-1)^\ell a^*_{\ell,-m,k}.
 \label{eq:mode_symmetry}
\end{align}
Together with the identity transformation $I$, they form the four-element group $G=\{I,E,P,EP\}$. Evaluating the unconstrained shared network $f_\theta$ on the four symmetry-transformed inputs $gx$ and forming the real-linear Reynolds average yields
\begin{equation}
 c(x)=\frac{s(\nu)}4\sum_{g\in G}D_g^{-1}f_\theta(gx),
 \label{eq:symmetry_projection}
\end{equation}
where
\begin{equation}
 s(\nu)=\left[1+(0.05/\nu)^4\right]^{-1/4}.
 \label{eq:nu-factor}
\end{equation}
This fixed, exchange-invariant prefactor approaches unity at comparable masses and imposes linear scaling with $\nu$ in the extreme-mass-ratio limit. The Reynolds average~\eqref{eq:symmetry_projection} guarantees exact equivariance $c(gx)=D_gc(x)$ for any network weights. In particular, Eq.~\eqref{eq:symmetry_projection} enforces $h_{\ell,-m}=(-1)^\ell h_{\ell m}^*$ for aligned-spin binaries ($Px=x$) and forces odd-$m$ modes to vanish for exchange-symmetric configurations ($Ex=x$), while retaining all 21 modes and reflection-asymmetric content for generic precessing binaries.

We define waveform alignment by a common time shift $\delta$ and a rigid azimuthal rotation $R_z(\alpha)$ by angle $\alpha$ about the reference $z$ axis across all modes:
\begin{equation}
 (\mathcal A_{\delta,\alpha}h)_{\ell m}(\tau)
 =e^{-im\alpha}h_{\ell m}(\tau-\delta),
 \label{eq:alignment}
\end{equation}
with the same transformation $(\mathcal A_{\delta,\alpha}\Psi_4)_{\ell m}(\tau)=e^{-im\alpha}\Psi_{4,\ell m}(\tau-\delta)$ applied to curvature. Minimizing a shape loss over $(\delta,\alpha)$ leaves the overall time origin and azimuthal phase free (Appendix~\ref{app:registration}); we therefore supplement the aligned NR shape loss with a soft unaligned anchor penalty at $\delta=\alpha=0$ on SXS waveforms and use fixed-azimuth supervision for the in-domain \Sur{} teachers (Appendix~\ref{app:gauge_sensitivity}).

\subsection{Strain training loss}\label{sec:training}

For training waveform $i$ with target strain modes $y^i_{\ell m}(\tau)$ sampled at $0.5M$ on $[-1500,100]M$ (3201 points) and binary observation mask $O_i(\tau)\in\{0,1\}$, let $Q_{i,W}=\sum_{\tau\in W}\sum_{\ell m}O_i(\tau)|y^i_{\ell m}(\tau)|^2$ denote the observed signal energy on interval $W$, and define the normalized squared residual on $W$ as
\begin{equation}
 R_{i,W}(\delta,\alpha)=
 \frac{\sum_{\ell m}\sum_{\tau\in W}O_i(\tau)
 |\mathcal A_{\delta,\alpha}h_{\ell m}-y^i_{\ell m}|^2}
 {Q_{i,W}}.
 \label{eq:window_loss}
\end{equation}
To weight the inspiral and merger regimes equally while supervising the early-inspiral and late-ringdown tails smoothly, we combine the core term $H^{\rm core}_i=(R_{i,[-1000,-160)}+R_{i,[-160,40]})/2$ (or $R_{i,[-160,40]}$ for merger-only targets) with an early-inspiral term $H^{\rm early}_i$ on $[-1500,-1000)$ (weight $0.1$, denominator floor $10^{-3}Q_{i,[-160,40]}$; omitted for merger-only targets) and a late-ringdown term $H^{\rm late}_i$ on $(40,100]$ whose target weight ($0.1$, denominator floor $0.005\,Q_{i,[-160,40]}$; zero for perturbative extreme-mass-ratio teachers) blends smoothly from the merger density at $\tau=40M$ to $\tau\geq48M$ via a $C^2$ quintic smoothstep (Appendix~\ref{app:gradient}), giving $H_i=H^{\rm core}_i+H^{\rm early}_i+H^{\rm late}_i$. Profiling over $\delta\in[-40,100]M$ and full azimuth $\alpha\in[0,2\pi)$ yields the balanced squared strain error
\begin{equation}
 \ell_i=\min_{\substack{-40\leq\delta\leq100\\\alpha\in[0,2\pi)}}H_i(\delta,\alpha),
 \label{eq:loss}
\end{equation}
with the mask $O_i(\tau)$ held fixed in target coordinates.

During joint pretraining, we also regularize physical first- and second-order finite differences on the $0.5M$ grid, $D_i(\delta,\alpha)$~\cite{Czarnecki2017Sobolev}, together with an inverse-amplitude-weighted ringdown tail loss $T_i(\delta,\alpha)$ on $(10,40]M$ (Appendix~\ref{app:gradient}). The joint-stage NR shape objective is
\begin{equation}
 U_i=\min_{\substack{-40\leq\delta\leq100\\\alpha\in[0,2\pi)}}
 \left[H_i+0.1\gamma_iD_i+0.1T_i\right],
 \label{eq:temporal_loss}
\end{equation}
where $\gamma_i\in\{0,1\}$ indicates whether target $i$ passes the finite-difference resolution check in Appendix~\ref{app:gradient}. Parameter gradients of the profiled objective are evaluated at the minimizing alignment $(\delta^*,\alpha^*)$ via the envelope theorem~\cite{MilgromSegal2002Envelope}. During SXS-only fine-tuning, only a late-ringdown second-derivative penalty $D^{\rm late}_i$ on $(39,99]M$ is kept with weight $\lambda_{\rm D2}$ to suppress residual post-merger oscillations.

To anchor the output time and azimuth to the input horizon triad, we include the unaligned core-window reference-frame penalty
\begin{equation}
 A_i=\begin{cases}H^{\rm core}_i(0,0),&i\text{ from SXS},\\0,&i\text{ from RIT},\end{cases}
 \label{eq:reference_anchor}
\end{equation}
with weight $\mu=0.01$ in both training stages. RIT waveforms are excluded from $A_i$ because transport from their sparse horizon snapshots introduces additional reference-frame uncertainty (Appendix~\ref{app:observations}).

\subsection{Distillation and NR fine-tuning}\label{sec:training-schedule}

Training combines NR simulations with synthetic teacher waveforms~\cite{Hinton2015Distillation}. The 41{,}444 synthetic targets comprise 33{,}000 \Sur{} waveforms, 8144 \SEOB{} waveforms (2000 aligned-spin and 6144 precessing), and 300 aligned-spin \textsc{BHPTNRSur2dq1e3} waveforms~\cite{Rink2024BHPTNRSur2dq1e3}; their parameter distributions and mode coverage are specified in Sec.~\ref{sec:data}.

For the 24{,}000 in-domain \Sur{} targets, the azimuth is held fixed at $\alpha=0$ and the relative time shift is prescribed by a training-only polynomial output-time map $\widehat d(x)$ fitted to 1282 SXS training simulations (Appendix~\ref{app:reference-inputs}). For all remaining synthetic teachers, $(\delta,\alpha)$ are optimized on $H_i$ within the teacher's modeled mode subspace, and the synthetic loss $V_i=H_i(\delta^*,\alpha^*)+0.1D_i(\delta^*,\alpha^*)$ is evaluated in that same subspace.

At optimizer update $s$, the joint pretraining objective is
\begin{align}
 \mathcal L_{\rm pre}(s)={}&a\langle w_i r_i(s)[U_i+\mu A_i]\rangle_{\rm NR}\notag\\
 &+b\,r(s)\langle V_i\rangle_{\rm teacher}+10^{-6}\|\theta\|_2^2,
 \label{eq:distillation}
\end{align}
where $a\simeq0.092$ and $b\simeq0.908$ weight the NR and synthetic minibatch losses in an approximate $1:10$ ratio ($a+b=1$), $w_i$ are the NR quality weights (Sec.~\ref{sec:data}), and angle brackets denote minibatch averages over 64 NR and 704 synthetic waveforms per joint update. The synthetic schedule multiplier $r(s)$ equals unity through update 24{,}000, decreases linearly to zero at update 32{,}000, and remains zero thereafter; the catalogue weight is $r_i(s)=1$ for SXS and $r_i(s)=r(s)$ for RIT.

We optimize the network using Adam~\cite{Kingma2015Adam} with gradient-norm clipping at 5, a 200-step warmup to $10^{-3}$ followed by a cosine decay to $10^{-5}$ at update 40{,}000, and an exponential moving average (EMA) of the weights (decay $0.999$). After update 32{,}000, when synthetic and RIT weights have annealed to zero, training draws minibatches of 32 from the 2038 SXS training waveforms alone and minimizes
\begin{equation}
 \mathcal L_{\rm ft}=a\left\langle w_i\left[\ell_i+\mu A_i+\lambda_{\rm D2}D^{\rm late}_i\right]\right\rangle_{\rm SXS},
 \label{eq:final_loss}
\end{equation}
with $\mu=0.01$ and $\lambda_{\rm D2}=0.02$ through update 52{,}000 (holding the learning rate at $10^{-5}$ past update 40{,}000), followed by 12{,}000 SXS-only updates to update 64{,}000 with $\lambda_{\rm D2}=0.40$. All reported results use the double-precision EMA weights at update 64{,}000, which achieve a validation SXS group-mean strain error of $0.0285$ across the 77 selected SXS validation groups. Network parameters and gradients are computed in single precision (using TensorFloat-32 in early pretraining and strict IEEE \texttt{float32} from update 28{,}000 through fine-tuning), while EMA weights and all NR alignment and loss reductions run in double precision.

%% file: sections/data.tex
\section{Data and evaluation}\label{sec:data}

\subsection{NR samples and reference conventions}

Table~\ref{tab:data} summarizes the SXS and RIT simulation partitions~\cite{Boyle2019SXS,Scheel2025SXS,Healy2022RIT}. Repeated resolutions and closely matching parameter configurations of the same binary are assigned to a common physical-system group. The training, validation, and evaluation partitions contain 2697 waveforms (2347 groups), 166 waveforms (150 groups), and 282 waveforms (266 groups), respectively, and are disjoint at the group level. Checkpoint selection omits one validation case (SXS:BBH:2633) whose two highest numerical resolutions diverge substantially before merger (Appendix~\ref{app:observations}), using the remaining 165 validation waveforms (149 groups: 81 SXS in 77 groups and 84 RIT in 77 groups).

\input{tables/current_data}

Since the synthetic teachers are themselves calibrated to NR waveforms, we track upstream NR membership across physical-system groups, including cross-catalogue matches (Appendix~\ref{app:teacher-exposure}). Identified \Sur{} fitting families, the \textsc{NRHybSur3dq8}~\cite{Varma2019NRHybSur3dq8} fitting families inherited through \textsc{BHPTNRSur2dq1e3} calibration, and identified \SEOB{} calibration or antisymmetric-parent families are assigned to training or validation. External \Sur{} $q=6$ test cases that were not used in fitting are eligible for evaluation. This removes the identified upstream waveform-fitting and parent-sample overlaps under our physical-system grouping policy, while the auxiliary reference-epoch regressions ($\widehat\Delta$ and $\widehat d$) and precessing \SEOB{} teacher radiation-axis alignment model are fitted exclusively on training simulations (Appendix~\ref{app:teacher-exposure}).

For SXS, the reference spins and orbital triad $(\boldsymbol n,\hat{\boldsymbol L}\times\boldsymbol n,\hat{\boldsymbol L})$ are interpolated from the continuous apparent-horizon trajectories at the last post-relaxation upward crossing of $M\Omega=\pi Mf_{\rm ref}$, with orbital frequency $\Omega=|\boldsymbol r\times\dot{\boldsymbol r}|/|\boldsymbol r|^2$ computed from the relative horizon separation $\boldsymbol r$ and its time derivative $\dot{\boldsymbol r}$. For RIT, which provides discrete horizon snapshots, reference spins and the orbital normal are transported to $Mf_{\rm ref}$ along a coupled quasicircular SEOB-based evolution, with waveform-clock and radiative-frame corrections (Appendix~\ref{app:observations}).

The target waveform time origin is defined independently of the spin-reference epoch. For native strain $h(u)$, let $P(u)=\sum_{\ell m}|\ddot h_{\ell m}(u)|^2$, evaluated on a $0.5M$ grid with a sixth-order finite-difference stencil. We set
\begin{align}
 t_0&=\frac{\int u\,w(u)\,du}{\int w(u)\,du},\\
 w(u)&=\left[P(u)\left(1-e^{-(u-u_0)^2/(2T^2)}\right)\right]^8,
\end{align}
where $u_0$ is the initial sample and $T=100M$ suppresses initial junk radiation. The dimensionless waveform time is $\tau=(u-t_0)/M$. Target waveforms are sampled on the $0.5M$ grid over $\tau\in[-1500,100]M$, with observation masks $O_i(\tau)$ excluding the initial $100M$ (with shorter margins for short waveforms) and any unmeasured intervals. SXS strain targets subtract the catalogue-added Bondi--Metzner--Sachs (BMS) energy-flux memory contribution as described in Appendix~\ref{app:observations}~\cite{Mitman2020memory}. For RIT strain targets, aligned-spin waveforms use the published strain and precessing waveforms use fixed-frequency double integration of the released $\Psi_4$ modes at $\omega_0=0.4(2\pi f_{22,\rm start})$, where $f_{22,\rm start}$ is the initial $(2,2)$-mode frequency, followed in both branches by Gaussian low-frequency drift removal (Appendix~\ref{app:observations}). Given that the RIT targets rely on transported reference frames and, for precessing systems, double-integrated $\Psi_4$, we report SXS and RIT evaluations separately and restrict curvature benchmarks to SXS and native BAM extractions.

\subsection{Synthetic supervision}

The synthetic surrogate teacher is the original \Sur{} model~\cite{Varma2019NRSur7dq4}; the updated \SurVTwo{} model is used only for evaluation. The 41{,}444 synthetic training waveforms, generated on $[-1620,160]M$ at $0.5M$ spacing and masked outside each generator's valid support (such as $t_{\rm sur}>100M$ for \Sur{}, corresponding to $\tau>90.5M$), comprise six subsets:
\begin{enumerate}
 \item 24{,}000 in-domain \Sur{} waveforms with $1\leq q\leq4$ and $|\boldsymbol\chi_{1,2}|\leq 0.8$;
 \item 8000 extrapolated \Sur{} waveforms with $4<q\leq6$ and $|\boldsymbol\chi_{1,2}|\leq 0.8$;
 \item 1000 aligned-spin \Sur{} waveforms with $1\leq q\leq4$ and $0.8<\max(|\chi_{1z}|,|\chi_{2z}|)\leq 0.95$;
 \item 2000 aligned-spin \SEOB{} waveforms with $q$ log-uniform in $[6,20]$ and $\chi_{1z,2z}$ uniform in $[-0.8,0.8]$;
 \item 6144 precessing \SEOB{} waveforms: 4608 with $q$ log-uniform in $[8,20]$ and 1536 in $[6,8]$. In each interval, half have isotropic spins of magnitude $[0,0.9]$ and half have primary spin magnitude $[0.5,0.9]$, primary tilt cosine $[-0.5,0.5]$, and secondary spin magnitude $[0,0.3]$;
 \item 300 aligned-spin \textsc{BHPTNRSur2dq1e3} waveforms~\cite{Rink2024BHPTNRSur2dq1e3} with $q$ log-uniform in $(20,1000]$, primary spin $\chi_{1z}\in[-0.5,0.8]$, and $\boldsymbol\chi_2=0$.
\end{enumerate}
The \Sur{} subsets (1)--(3) sample $q$ uniformly. The \SEOB{} subsets (4)--(5) extend multipolar pretraining to $q\in[6,20]$ and high precessing spins where direct surrogate coverage is unavailable.

A waveform generator's nominal reference frequency does not necessarily correspond to the same time relative to merger as the NR horizon orbital frequency. For \Sur{}, we therefore hold the prescribed spin vectors fixed and shift its internal reference-time map $t_{\rm map}^{\rm Sur}(x)$ by
\begin{equation}
 t_{\rm ref}^{\rm Sur}(x)=t_{\rm map}^{\rm Sur}(x)+\widehat\Delta(x),
 \label{eq:sur_tref}
\end{equation}
where $\widehat\Delta(x)$ is a 60-feature cubic ridge polynomial (ridge penalty $0.01$, even under reversal of the in-plane spin components) fitted to 798 SXS training simulations (Appendix~\ref{app:reference-inputs}). \SEOB{} teachers are generated at $f_{\rm ref}=20\,\mathrm{Hz}$ and $M=120\,M_\odot$ from a $10\,\mathrm{Hz}$ start frequency with $0.5M$ sampling using \texttt{pyseobnr} 0.3.6 with the $(2,2)$, $(3,3)$, and $(4,4)$ antisymmetric mode corrections enabled. For the 6144 precessing \SEOB{} training teachers, we apply a time-independent, parameter-dependent active $SO(3)$ radiation-axis rotation fitted to 43 SXS training simulations ($q\in[5,6]$) to reduce high-$q$ frame misalignment relative to SXS NR, applying the covariant rotation to both the 21-mode waveforms and the rank-12 coprecessing observation operator prior to training (Appendix~\ref{app:observations}). Both the aligned-spin \SEOB{} teachers and the \SEOB{} evaluation comparator use unmodified \texttt{pyseobnr} 0.3.6.

\Sur{} provides all 21 modes across its supported time span, \textsc{BHPTNRSur2dq1e3} provides 16 modes [excluding $m=0$ and $(4,\pm1)$], and aligned-spin \SEOB{} provides the 12 modes $(2,\pm1)$, $(2,\pm2)$, $(3,\pm2)$, $(3,\pm3)$, $(4,\pm3)$, and $(4,\pm4)$, with unmodeled modes masked out of the loss. For precessing \SEOB{}, the same 12 modes are defined in the aligned coprecessing frame and span a rank-12 rotating subspace of the 21 inertial modes; a time-dependent projection operator $\mathsf P_i(\tau)$, constructed from the covariant coprecessing frame and Wigner $D$-matrices, projects both prediction and target into this subspace prior to evaluating strain and finite-difference residuals.

In the training objective~\eqref{eq:distillation}, each physical-system group carries unit total weight $w_i$, distributed among repeated simulations of the same system according to their resolution quality.

\subsection{Reference inputs for comparison models}\label{sec:input-construction}

At evaluation, all models receive the same seven physical input parameters $x=(q,\boldsymbol\chi_1,\boldsymbol\chi_2)$ on the $0.5M$ grid $[-1620,160]M$ and are scored on $[-1500,100]M$. The EOB baseline \SEOB{} (\texttt{pyseobnr} 0.3.6; denoted SEOB in tables) is evaluated with its antisymmetric $(2,2)$, $(3,3)$, and $(4,4)$ mode corrections enabled~\cite{RamosBuades2023SEOBNRv5PHM,Estelles2025Asym} at $f_{\rm ref}=20\,\mathrm{Hz}$ and $M=120\,M_\odot$, starting from $10\,\mathrm{Hz}$ at $0.5M$ spacing and centered $10M$ after its total-strain amplitude peak prior to alignment optimization. The time-domain phenomenological baseline \TPHM{}~\cite{Estelles2022TPHM} (\texttt{LALSuite} 7.26; denoted TPHM in tables) is evaluated for all 21 signed $\ell\leq 4$ modes at the same $f_{\rm ref}=20\,\mathrm{Hz}$, $f_{\rm start}=10\,\mathrm{Hz}$, and $M=120\,M_\odot$ with default precession settings, resampled to $0.5M$ spacing and centered $10M$ after its total-strain amplitude peak. \SurVTwo{}~\cite{Ravishankar2026NRSur7dq4v2} uses all 21 signed $\ell\leq4$ modes without the fast-evaluation preset, initializing the input spin vectors at the predicted reference event $\widehat T(x)=t_{\rm map}^{\rm Sur}(x)+\widehat\Delta(x)$ from Eq.~\eqref{eq:sur_tref} (Appendix~\ref{app:reference-inputs}). Since \SurVTwo{} terminates natively at $t_{\rm sur}=100M$ ($\tau=90.5M$), it is evaluated natively on $[-1620,90.5]M$ and zero-padded over $(90.5,160]M$; Appendix~\ref{app:teacher-exposure} shows that restricting both the Transformer and \SurVTwo{} to the unpadded $[-1500,50]M$ interval leaves all comparisons essentially unchanged. Time translation and rigid azimuthal rotation are optimized using the same scoring prescription for every model~\cite{MacUilliam2024Survey}.

For this comparison, we restrict \SurVTwo{} extrapolation to $q\leq6$, allowing a $0.01$ tolerance in the catalogue mass ratio. Four SXS and six RIT evaluation waveforms fall outside this range. Direct four-model rankings use the 105 supported SXS waveforms in 102 groups; full-sample results and unsupported counts are reported separately. We also report the common calibration-domain subset ($q\leq4.01$, $|\boldsymbol\chi_{1,2}|\leq0.8$).

\begin{table*}[t]
\centering
\caption{Model accuracy on $[-1500,100]M$ for common SXS support (105 waveforms, 102 physical systems), with full supported-sample statistics shown separately (109 waveforms, 106 systems for the Transformer, \SEOB{}, and \TPHM{}; 105/102 for \SurVTwo{}). Mismatch statistics use $M=120\,M_\odot$. Means weight systems equally; P95 is a waveform quantile.}\label{tab:product-v6-headline}
\input{tables/product_v6_headline}
\end{table*}

\begin{figure*}[tp]
\centering\includegraphics[width=\textwidth]{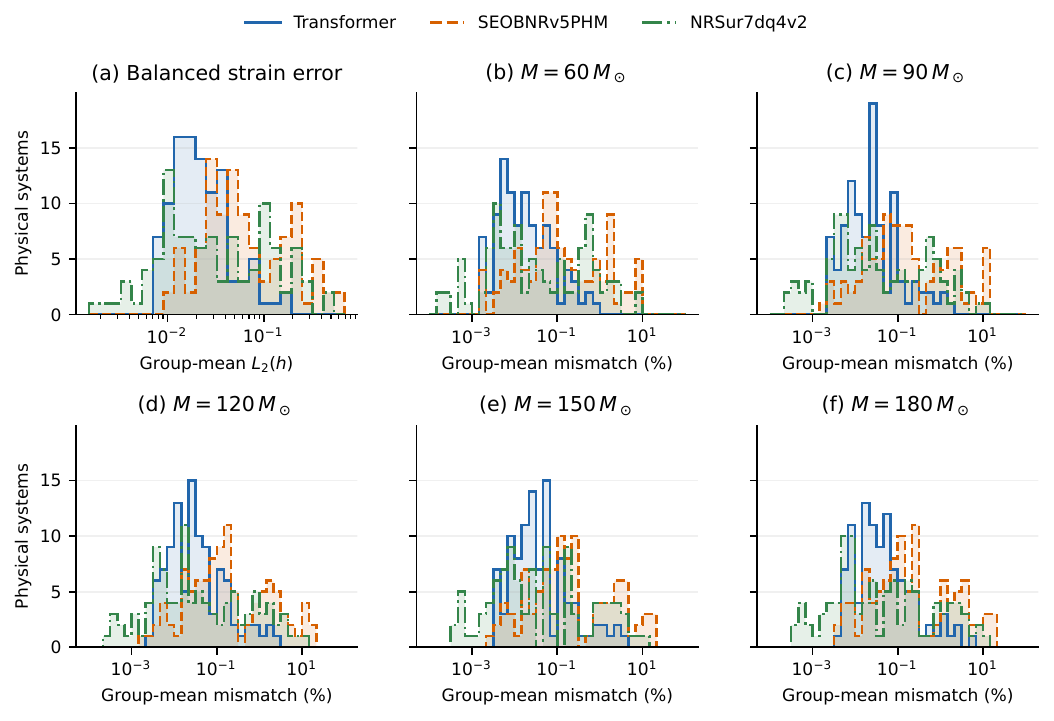}
\caption{Error distributions on $[-1500,100]M$ for the common sample of 105 SXS evaluation waveforms in 102 physical systems. Panel (a) shows the balanced strain error $L_2(h)$; panels (b)--(f) show detector-weighted mismatches at $M=60,90,120,150,180\,M_\odot$, respectively. Each entry is a physical-system mean. The histograms omit \TPHM{} for visual clarity; its results are included in Table~\ref{tab:product-v6-headline} and Fig.~\ref{fig:mass-mismatch}. All mismatch panels use the same bins, the $[-1500,100]M$ window, and the Advanced LIGO design PSD over 20--512 Hz.}
\label{fig:diagnostic}\label{fig:mismatch}
\end{figure*}

\begin{figure*}[tp]
\centering\includegraphics[width=\textwidth]{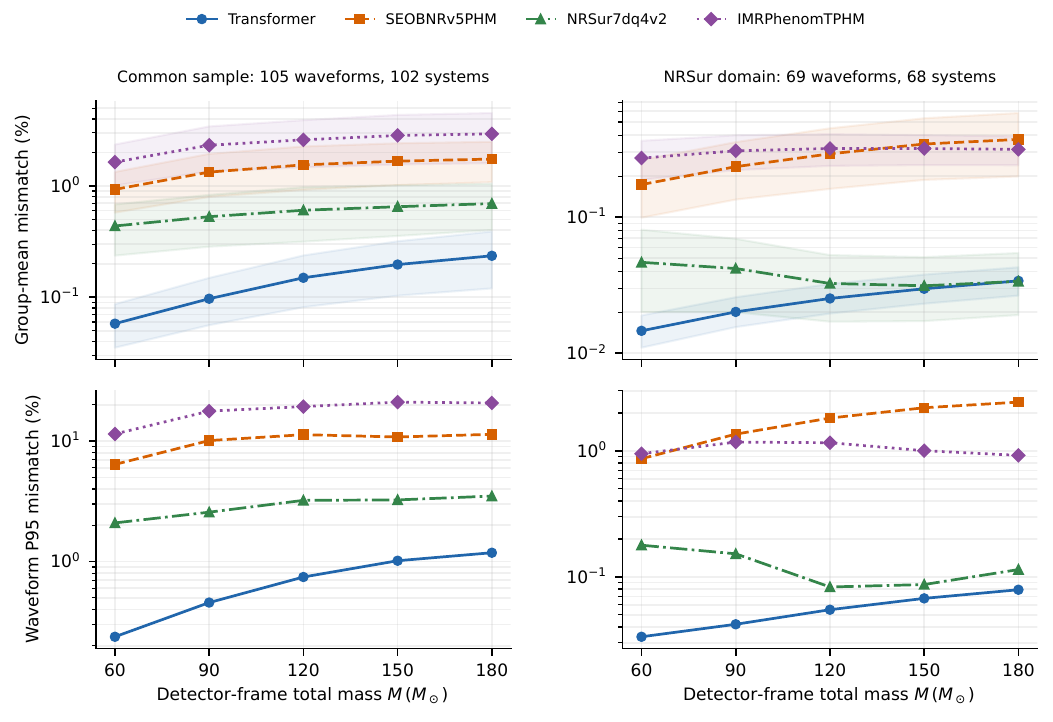}
\caption{Mass dependence of detector-weighted mismatch on $[-1500,100]M$ for the four comparison models. Left: the common 105-waveform, 102-system SXS sample. Right: the 69 waveforms in 68 systems within the \SurVTwo{} calibration domain ($q\leq4.01$, $\chi\leq0.8$). Top: physical-system means, with pointwise 95\% group-bootstrap intervals. Bottom: waveform-level P95. The dimensionless inputs and $[-1500,100]M$ window are fixed across detector-frame total masses.}
\label{fig:mass-mismatch}
\end{figure*}

\begin{table*}[tp]
\centering
\caption{Group-mean balanced strain error $L_2(h)$ on $[-1500,100]M$ by SXS evaluation stratum, with $\chi=\max(|\boldsymbol\chi_1|,|\boldsymbol\chi_2|)$.}\label{tab:product-v6-strata}
\input{tables/product_v6_strata}
\end{table*}

\subsection{Evaluation metrics}\label{sec:metric}

Our primary time-domain accuracy metric is the balanced root-mean-square strain error
\begin{equation}
 L_2(h)=\sqrt{\frac{1}{2}\left(R_{i,[-1500,-160)}+R_{i,[-160,100]}\right)},
\end{equation}
evaluated over all 21 modes on $[-1500,100]M$ ($1M$ evaluation lattice) with full translation and azimuthal profiling ($\delta\in[-40,100]M$, $\alpha\in[0,2\pi)$) using identical alignment bounds across all models.

To quantify accuracy in a gravitational-wave detector band~\cite{Lindblom2008Accuracy}, we compute the noise-weighted inner product for complex projected strain $a(t)$ and $b(t)$ at detector-frame total masses $M=60,90,120,150,180\,M_\odot$,
\begin{equation}
 \langle a,b\rangle_S=\int_{20\leq|f|\leq512\,{\rm Hz}}
 \frac{\widetilde a(f)\widetilde b(f)^*}{S_n(|f|)}\,df,
\end{equation}
where $\widetilde a(f)$ and $\widetilde b(f)$ are the Fourier transforms of $a(t)$ and $b(t)$, $S_n(f)$ is the Advanced LIGO zero-detuned high-power power spectral density (PSD)~\cite{LIGOZDHP}, and $\|a\|_S=\sqrt{\langle a,a\rangle_S}$. At each of seven observer inclinations $\iota=0,\pi/6,\ldots,\pi$ (with reference azimuth $\varphi=0$), let $\mathcal A_{\delta,\alpha}h_\iota(t)=\sum_{\ell m}(\mathcal A_{\delta,\alpha}h)_{\ell m}(t/M)\,{}_{-2}Y_{\ell m}(\iota,0)$ and $h^i_\iota(t)=\sum_{\ell m}y^i_{\ell m}(t/M)\,{}_{-2}Y_{\ell m}(\iota,0)$ denote the aligned model and target complex polarizations; the inclination mismatch is
\begin{equation}
 \mathcal M_\iota=1-\max_{\substack{-40\leq\delta\leq100\\\alpha\in[0,2\pi)}}
 \frac{|\langle\mathcal A_{\delta,\alpha}h_\iota,h^i_\iota\rangle_S|}
 {\|\mathcal A_{\delta,\alpha}h_\iota\|_S\|h^i_\iota\|_S},
 \label{eq:mm}
\end{equation}
where the absolute value maximizes over a constant polarization rotation $h\mapsto e^{-2i\psi}h$ (distinct from the source azimuth $\alpha$, which acts mode-by-mode as $e^{-im\alpha}$). Cosine tapers of $10\text{ ms}$ duration are applied to the masked time-domain window before an 8192-point two-sided fast Fourier transform (FFT), with $\|\mathcal A_{\delta,\alpha}h_\iota\|_S$ recomputed at each $(\delta,\alpha)$. Each waveform's mismatch $\mathcal M$ is the median of $\mathcal M_\iota$ over the seven inclinations, quoted as a percentage $100\mathcal M$.

The mass scan rescales the same dimensionless waveforms and holds $Mf_{\rm ref}$ fixed, so that $f_{\rm ref}=20\,\mathrm{Hz}\,(120\,M_\odot/M)$. The detector weights, physical taper, and optimal alignment are evaluated separately at each mass on the $[-1500,100]M$ window ($0.473\,\mathrm{s}$ at $60\,M_\odot$). Unless otherwise stated, tabulated mismatch values use $120\,M_\odot$.

For SXS targets passing the interpolation checks in Appendix~\ref{app:gradient}, we also evaluate the relative 21-mode curvature error $L_2(\Psi_4)=\sqrt{C_i(\delta^*,\alpha^*)}$ on $[-160,40]M$ at $0.5M$ spacing (with $C_i$ defined in Appendix~\ref{app:gradient}) using each model's $[-1500,100]M$ strain-fitted alignment $(\delta^*,\alpha^*)$, where the target $\psi^i_{\ell m}=-\partial_\tau^2y^i_{\ell m}$ is obtained from a quintic spline interpolant of the native SXS strain.

In all tables, summary means first average waveform scores within each physical-system group and then weight groups equally, whereas medians and 95th percentiles (P95) are computed over individual waveforms. Paired bootstrap confidence intervals resample common physical-system groups.

%% file: tables/current_data.tex
\begin{table}[t]
\centering
\caption{NR data partition. Physical-system groups are counted across catalogues. The evaluation subsets contain 106 SXS groups and 163 RIT groups; three are shared, giving $106+163-3=266$ distinct systems. Validation contains 166 waveforms (150 groups), with 165 waveforms (149 groups: 81 SXS in 77 groups and 84 RIT in 77 groups, five shared) used for checkpoint selection after excluding SXS:BBH:2633.}\label{tab:data}
\begin{tabular}{lrrrr}
\toprule
Set & SXS & RIT & Total & Groups \\
\midrule
Training & 2038 & 659 & 2697 & 2347 \\
Validation (tot./sel.) & 82/81 & 84/84 & 166/165 & 150/149 \\
Evaluation & 109 & 173 & 282 & 266 \\
\bottomrule
\end{tabular}
\end{table}

%% file: tables/product_v6_headline.tex
\begin{tabular}{lrrrrr}
\toprule
Model & Waveforms/groups & $L_2(h)$ & $L_2(\Psi_4)$ & Mean $\mathcal M$ (\%) & P95 (\%) \\
\midrule
\multicolumn{6}{l}{Common support: 105 waveforms / 102 groups} \\
Transformer & 105/102 & 0.0301 & 0.1487 & 0.1496 & 0.7446 \\
\SEOB{} & 105/102 & 0.1124 & 0.3262 & 1.5475 & 11.3127 \\
\SurVTwo{} & 105/102 & 0.0679 & 0.2155 & 0.6049 & 3.2233 \\
\TPHM{} & 105/102 & 0.1546 & 0.3752 & 2.5983 & 19.3484 \\
\midrule
\multicolumn{6}{l}{Full SXS panel: 109 waveforms / 106 groups} \\
Transformer & 109/106 & 0.0305 & 0.1497 & 0.1536 & 0.7870 \\
\SEOB{} & 109/106 & 0.1123 & 0.3249 & 1.5886 & 11.2267 \\
\TPHM{} & 109/106 & 0.1547 & 0.3782 & 2.6427 & 19.0632 \\
\bottomrule
\end{tabular}

%% file: tables/product_v6_strata.tex
\begin{tabular}{llrrrrr}
\toprule
Spin & Mass-ratio / spin bin & Waveforms/groups & Transformer & SEOB & NRSur & TPHM \\
\midrule
Aligned & $q\leq4$, $\chi\leq0.8$ & 23/23 & 0.0143 & 0.0274 & 0.0118 & 0.0341 \\
 & $q\leq4$, $\chi>0.8$ & 0/0 & -- & -- & -- & -- \\
 & $4<q\leq6$ & 1/1 & 0.0104 & 0.0178 & 0.0082 & 0.0368 \\
 & $6<q\leq8$ & 1/1 & 0.0164 & 0.0200 & -- & 0.0469 \\
 & $q>8$ & 2/2 & 0.0178 & 0.0214 & -- & 0.0496 \\
\midrule
Precessing & $q\leq4$, $\chi\leq0.8$ & 46/45 & 0.0205 & 0.0646 & 0.0303 & 0.0954 \\
 & $q\leq4$, $\chi>0.8$ & 10/9 & 0.0412 & 0.2035 & 0.0971 & 0.2033 \\
 & $4<q\leq6$ & 25/24 & 0.0599 & 0.2531 & 0.1838 & 0.3676 \\
 & $6<q\leq8$ & 1/1 & 0.1085 & 0.3768 & -- & 0.4830 \\
 & $q>8$ & 0/0 & -- & -- & -- & -- \\
\bottomrule
\end{tabular}

%% file: sections/results.tex
\section{Evaluation results}\label{sec:results}

\subsection{Strain accuracy}\label{sec:confirmation}

We evaluate the selected model on 282 NR simulations in 266 physical-system groups over the $[-1500,100]M$ comparison window, beginning with the \ProductRows\ SXS waveforms in \ProductGroups\ groups. Table~\ref{tab:product-v6-headline} summarizes the balanced strain error $L_2(h)$, strain-derived curvature error $L_2(\Psi_4)$, and finite-window detector mismatch $\mathcal M$ at $120\,M_\odot$ for the Transformer, \SEOB{}, \SurVTwo{}, and \TPHM{}. Direct four-model comparisons use the 105 SXS waveforms in 102 systems supported by all four models; full-sample statistics are listed separately. The Transformer's full-sample group-mean strain error is \ProductStrainMean\ and its mean mismatch is \ProductMismatchMean\%. On common support, its group-mean errors are \CommonProductStrainMean\ and \CommonProductMismatchMean\%, below \SurVTwo{} (\CommonNRSurStrainMean, \CommonNRSurMismatchMean\%), \SEOB{} (\CommonSEOBStrainMean, \CommonSEOBMismatchMean\%), and \TPHM{} (\CommonTPHMStrainMean, \CommonTPHMMismatchMean\%). For reference on the shorter $[-1000,40]M$ core interval, the Transformer yields full-SXS group means $L_2(h)=\CoreRefSXSStrain$ and $\mathcal M(120\,M_\odot)=\CoreRefSXSMismatch\%$ ($\CoreRefCommonStrain$ on common support). The $[-1500,100]M$ error distributions are shown in Fig.~\ref{fig:diagnostic}.

Figure~\ref{fig:mass-mismatch} shows the mass dependence of mismatch means and P95 values on the common 105-waveform sample over $[-1500,100]M$. From $60$ to $180\,M_\odot$, the Transformer's group-mean mismatch rises gently from $\VSevenCommonSixty\%$ to $\VSevenCommonOneEighty\%$, compared with $\SurCommonSixty\%$ to $\SurCommonOneEighty\%$ for \SurVTwo{}, $\SeobCommonSixty\%$ to $\SeobCommonOneEighty\%$ for \SEOB{}, and $\TphmCommonSixty\%$ to $\TphmCommonOneEighty\%$ for \TPHM{}. Both its mean and its P95 mismatch remain below those of all three comparators at every tested total mass, with 95\% group-bootstrap confidence intervals for the paired differences excluding zero across the entire mass scan. Evaluated on the $0.5M$ SXS targets, the Transformer achieves mean relative $L_2$ errors of $\EarlyExtStrain$ on the early inspiral $[-1500,-1000)M$, $\FullDenseStrain$ across the full $[-1500,100]M$ window, and $\LateExtStrain$ in the post-$40M$ ringdown tail $(40,100]M$ (compared with $0.461$ for \SEOB{} and $0.434$ for \TPHM{}).

\begin{table*}[tp]
\centering
\caption{Paired group-mean differences (Transformer minus comparator) on $[-1500,100]M$ over common SXS physical systems, with 95\% bootstrap confidence intervals from 10{,}000 group resamples. Mismatch differences $\Delta\mathcal{M}$ are in percentage points; strain and curvature differences are dimensionless.}\label{tab:product-v6-paired}
\input{tables/product_v6_paired}
\end{table*}
\begin{figure*}[tp]
\centering\includegraphics[width=0.98\textwidth]{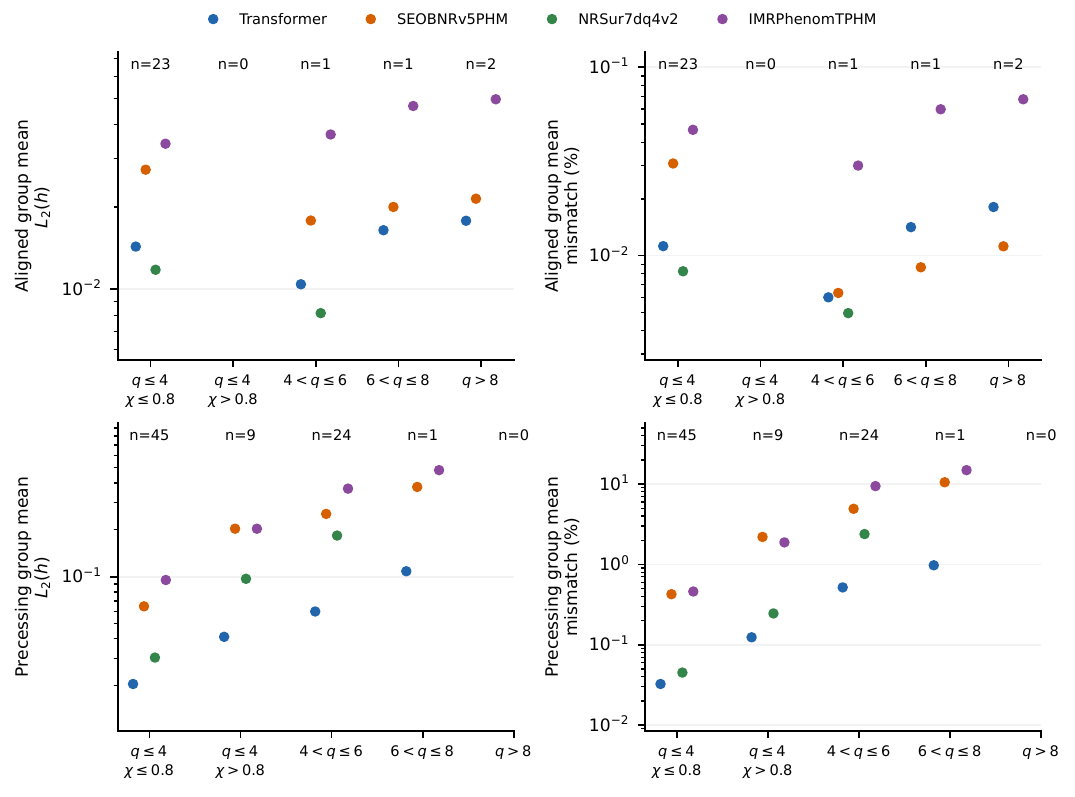}
\caption{Physical-system mean strain errors and detector mismatches on $[-1500,100]M$ by mass-ratio and spin-alignment stratum on the SXS evaluation set, labeled by the number of physical systems $n$ in each bin. The bins match Table~\ref{tab:product-v6-strata}, including the $0.01$ tolerance at mass-ratio boundaries.}
\label{fig:surrogate}
\end{figure*}
\begin{figure*}[tp]
\centering\includegraphics[width=\textwidth]{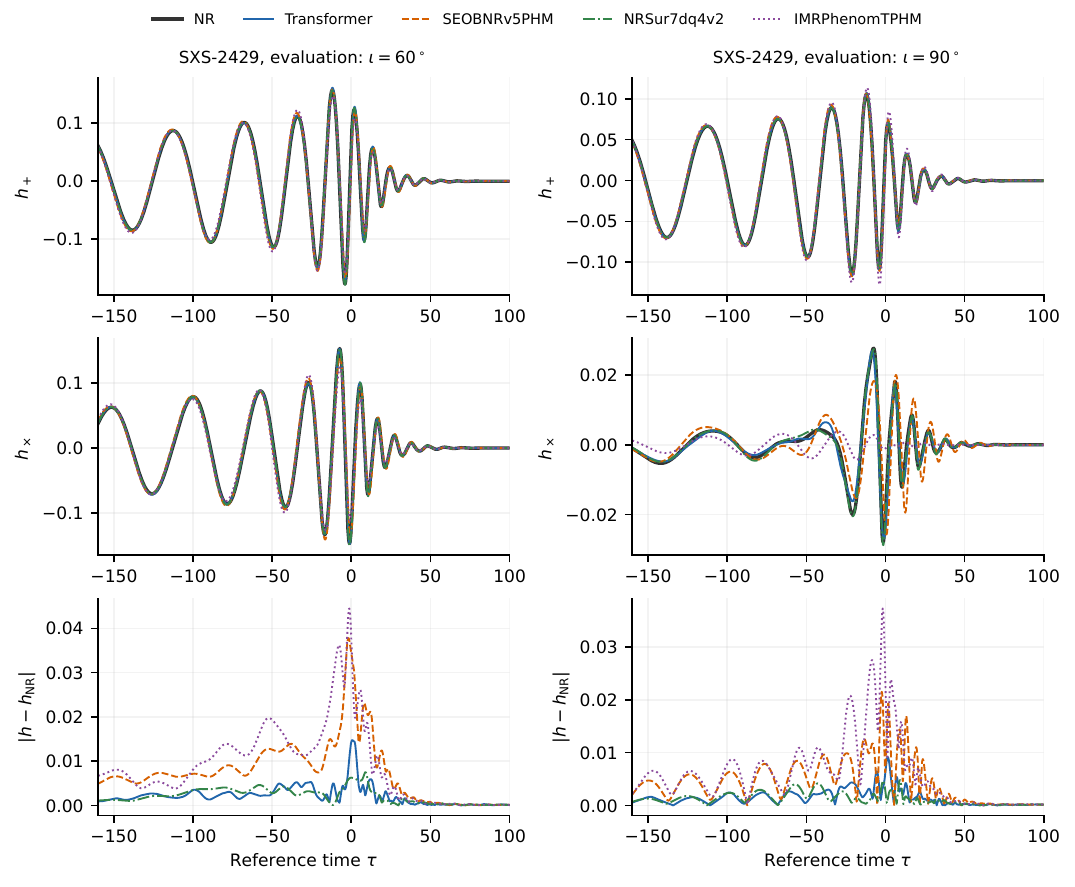}
\caption{Polarizations $h_+$, $h_\times$, and complex strain residuals $|h-h_{\rm NR}|$ for the precessing evaluation system SXS:BBH:2429 ($q\simeq1$, $\chi\simeq0.60$) at observer inclinations $\iota=60^\circ$ and $90^\circ$, using each model's $[-1500,100]M$ all-mode alignment.}
\label{fig:confirmation-waveform}
\end{figure*}
\begin{figure*}[tp]
\centering\includegraphics[width=\textwidth]{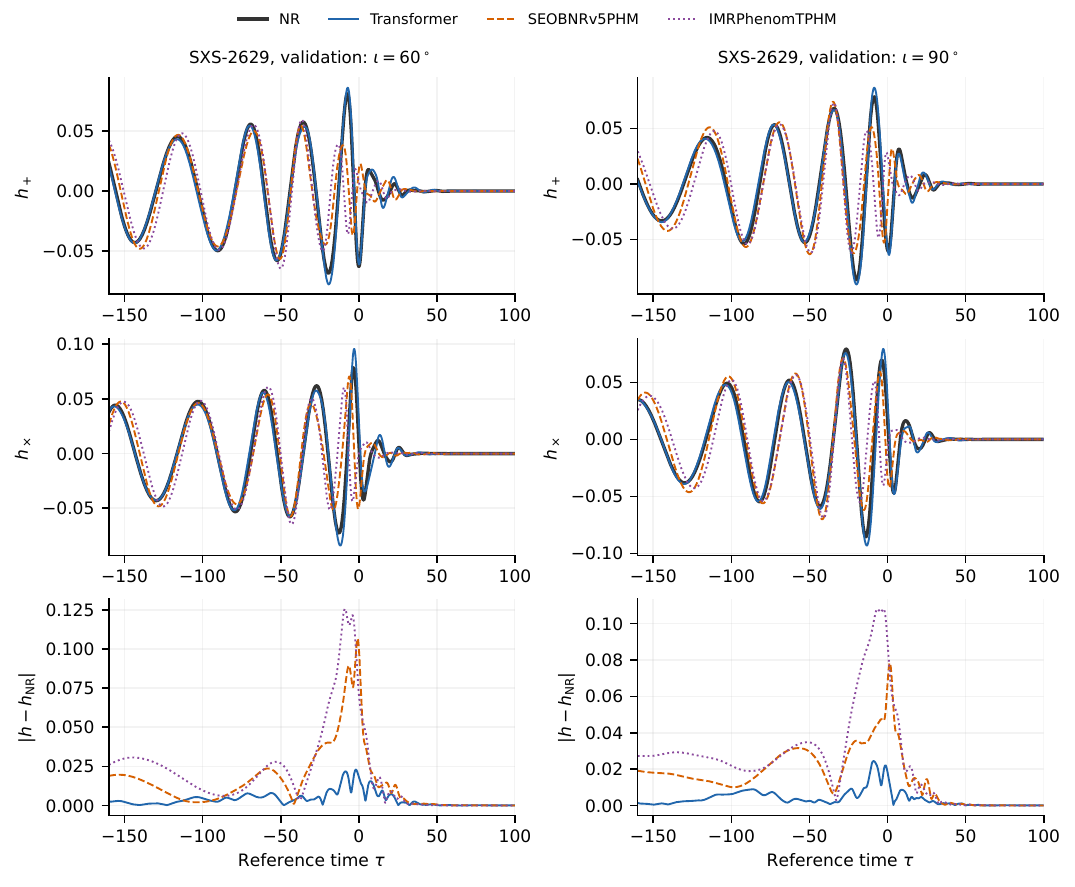}
\caption{Polarizations and complex strain residuals for the $q\simeq10$ precessing validation system SXS:BBH:2629 (\SurVTwo{} does not support $q>6$), using each model's $[-1500,100]M$ all-mode alignment.}
\label{fig:q10}
\end{figure*}

To examine performance across parameter space, we stratify the SXS evaluation sample by aligned-spin versus precessing classification, mass ratio (with a $0.01$ boundary tolerance), and spin magnitude $\chi=\max(|\boldsymbol\chi_1|,|\boldsymbol\chi_2|)$. Table~\ref{tab:product-v6-strata} and Fig.~\ref{fig:surrogate} show the resulting group means, and Table~\ref{tab:product-v6-paired} lists paired group-bootstrap differences on common support. Across precessing binaries---where the evaluation set contains 45 in-domain systems, 9 high-spin systems, 24 systems in $4<q\leq 6$, and one in $6<q\leq 8$---the Transformer improves uniformly over all three comparators in every populated stratum. In compiled \texttt{float64} evaluation after warm-up, generating all 21 modes on $[-1620,160]M$ at $0.5M$ spacing (3561 samples, including four-way symmetry projection and time-domain spline evaluation) reaches $\CompiledGPUThroughput\text{ waveforms/s}$ ($\CompiledGPUAmortizedMS\text{ ms}$ per waveform; batch of 1024) on an NVIDIA A100-SXM4-40GB GPU and $\CompiledCPUThroughput\text{ waveforms/s}$ ($\CompiledCPUAmortizedMS\text{ ms}$ per waveform; batch of \CompiledCPUBatch{} on \CompiledCPUThreads{} threads) on an AMD EPYC 7773X CPU (Appendix~\ref{app:runtime}).

\subsection{Surrogate-domain and catalogue comparisons}\label{sec:surrogate-comparison}

Table~\ref{tab:product-v6-inbox} restricts the comparison to the 69 SXS evaluation waveforms (68 systems) inside the \SurVTwo{} calibration domain ($q\leq4.01$, $\chi\leq0.8$). Even within this calibrated box, the Transformer attains a lower group-mean strain error ($\VSevenInboxStrain$ versus $\SurInboxStrain$, paired difference $\InboxStrainDiff$ with $95\%$ interval $[\InboxStrainDiffLo,\InboxStrainDiffHi]$), a lower P95 mismatch at $120\,M_\odot$ ($\VSevenInboxP\%$ versus $\SurInboxP\%$), and lower group-mean mismatches at $60$ and $90\,M_\odot$ where the earlier inspiral carries higher weight. Mean mismatches at $120\,M_\odot$ ($\VSevenInboxMismatch\%$ versus $\SurInboxMismatch\%$) and $180\,M_\odot$ ($\VSevenInboxHighMass\%$ versus $\SurInboxHighMass\%$; Fig.~\ref{fig:mass-mismatch}, right) are statistically comparable within 95\% bootstrap intervals, while \SurVTwo{} retains the lower median mismatch ($\SurInboxMedian\%$ versus $\VSevenInboxMedian\%$) and curvature error ($\SurInboxCurvature$ versus $\VSevenInboxCurvature$). Table~\ref{tab:nrsur-domain-sensitivity} verifies that restricting the evaluation end time from $100M$ to $50M$ so \SurVTwo{} requires no post-$90.5M$ zero-padding shifts $120\,M_\odot$ means only slightly ($\VSevenInboxMismatch\%\to\UnpaddedInboxLWMOneTwenty\%$ and $\SurInboxMismatch\%\to\UnpaddedInboxSurMOneTwenty\%$ in box; $\CommonProductMismatchMean\%\to\UnpaddedCommonLWMOneTwenty\%$ and $\CommonNRSurMismatchMean\%\to\UnpaddedCommonSurMOneTwenty\%$ on common support).

\begin{table}[t]
\centering
\caption{Comparison on $[-1500,100]M$ over the 69 SXS evaluation waveforms (68 physical systems) within the \SurVTwo{} calibration domain ($q\leq4$, $\chi\leq0.8$). Means are taken over physical systems; median and P95 mismatches are computed over individual waveforms.}\label{tab:product-v6-inbox}
\resizebox{\columnwidth}{!}{\input{tables/product_v6_inbox}}
\end{table}

Table~\ref{tab:product-v6-rit} reports balanced strain errors on the 173 RIT evaluation waveforms in 163 physical systems over $[-1500,100]M$. The Transformer gives $\VSevenRitAll$ overall, $\VSevenRitAligned$ for aligned-spin binaries, and $\VSevenRitPrecessing$ for precessing binaries. On the 167-waveform (157-group) common RIT subset, the Transformer ($\VSevenRitCommon$) matches \SurVTwo{} ($\SurRitCommon$, paired difference consistent with zero) and substantially improves upon \SEOB{} ($0.1082$) and \TPHM{} ($0.1184$).

\begin{table}[t]
\centering
\caption{Transformer balanced strain errors $L_2(h)$ on $[-1500,100]M$ over the 173 reconstructed RIT evaluation waveforms (163 physical systems). Means are taken over physical groups; medians and P95 are computed over individual waveforms.}\label{tab:product-v6-rit}
\input{tables/product_v6_rit}
\end{table}

\begin{table*}[t]
\centering
\caption{Reference-frame diagnostics for the Transformer on the SXS evaluation set. Waveforms are aligned by their independently estimated total-amplitude peaks on $[-160,20]M$ at fixed azimuth ($\alpha=0$) and after a single rigid $R_z(\alpha)$ rotation. Strain errors are physical-system means; $|\Delta\alpha|$ and $|\Delta t_{\rm peak}|/M$ are waveform medians.}\label{tab:product-v6-reference}
\input{tables/product_v6_reference}
\end{table*}

\begin{figure*}[t]
\centering\includegraphics[width=\textwidth]{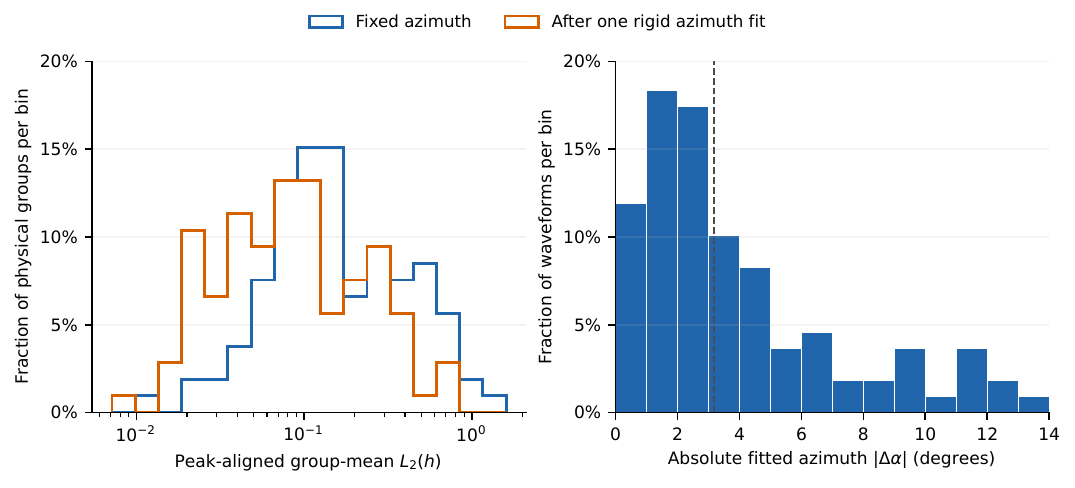}
\caption{Reference-frame diagnostics on the SXS evaluation set. Left: distributions of physical-system mean strain errors on $[-160,20]M$ at peak alignment with fixed azimuth ($\alpha=0$) and after one rigid azimuthal rotation. Right: distribution of absolute fitted azimuthal angles $|\Delta\alpha|$ in $1^\circ$ bins over $0$--$14^\circ$ (the dashed line marks the median $\ProductMedianAngleDeg^\circ$); 12 of the 109 waveforms lie above the displayed range, and the normalization retains all waveforms.}
\label{fig:reference-frame}
\end{figure*}

\begin{figure}[tbp]
\centering\includegraphics[width=\columnwidth]{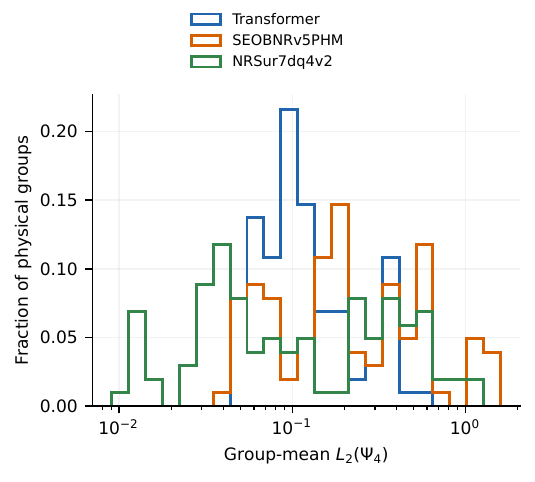}
\caption{Distributions of physical-group mean curvature error $L_2(\Psi_4)$ on $[-160,40]M$ over the common 105-waveform SXS sample (102 groups), evaluated at each model's $[-1500,100]M$ strain-fitted alignment. \TPHM{} is omitted for visual clarity; its statistics are given in Table~\ref{tab:product-v6-headline}.}
\label{fig:curvature}
\end{figure}

\subsection{Waveform examples}

Figure~\ref{fig:confirmation-waveform} compares the predicted polarizations $h_+$ and $h_\times$ and the complex strain residual $|h-h_{\rm NR}|$ over $[-160,100]M$ for the in-domain precessing evaluation binary SXS:BBH:2429 ($q\simeq1$, $\chi\simeq0.60$), where the $[-1500,100]M$ balanced strain errors are $\ExTwoFourTwoNineLW$ (Transformer), $\ExTwoFourTwoNineSur$ (\SurVTwo{}), $\ExTwoFourTwoNineSEOB$ (\SEOB{}), and $\ExTwoFourTwoNineTPHM$ (\TPHM{}). Figure~\ref{fig:q10} shows the $q\simeq10$ precessing validation binary SXS:BBH:2629, where the Transformer achieves a balanced strain error of $\ExTwoSixTwoNineLW$ compared with $\ExTwoSixTwoNineSEOB$ for \SEOB{} and $\ExTwoSixTwoNineTPHM$ for \TPHM{}. Both figures use each model's all-mode alignment $(\delta^*,\alpha^*)$ on $[-1500,100]M$ at observer azimuth $\varphi=0$.

\subsection{Reference-frame and derivative diagnostics}\label{sec:curvature-results}

To quantify reference-frame preservation without alignment profiling, we align each predicted SXS evaluation waveform to its NR target by the difference of their independently estimated total-amplitude peak times and evaluate the 21-mode relative strain error on $[-160,20]M$ at fixed azimuth ($\alpha=0$), as well as unaligned across the full $[-1500,100]M$ window ($\ProductFixedReferenceLTwo$). Over all \ProductGroups\ SXS evaluation groups on the peak-aligned $[-160,20]M$ diagnostic (Table~\ref{tab:product-v6-reference} and Fig.~\ref{fig:reference-frame}), the mean fixed-azimuth strain error is \ProductFixedLTwo; optimizing a single rigid azimuthal rotation $\alpha$ at the same peak time reduces this error to \ProductPeakRotatedLTwo. Over all \ProductRows\ waveforms, the median absolute azimuthal offset is $|\Delta\alpha|=\ProductMedianAngleDeg^\circ$ and the median absolute peak-time offset is $|\Delta t_{\rm peak}|=\ProductMedianPeakShift M$.

On the SXS evaluation targets, the model's group-mean analytic curvature error at the $[-1500,100]M$ strain-fitted alignment is \ProductCurvatureMean\ (Table~\ref{tab:product-v6-headline}); Fig.~\ref{fig:curvature} compares the corresponding curvature-error distributions among the models.

\begin{figure*}[tp]
\centering\includegraphics[width=0.76\textwidth]{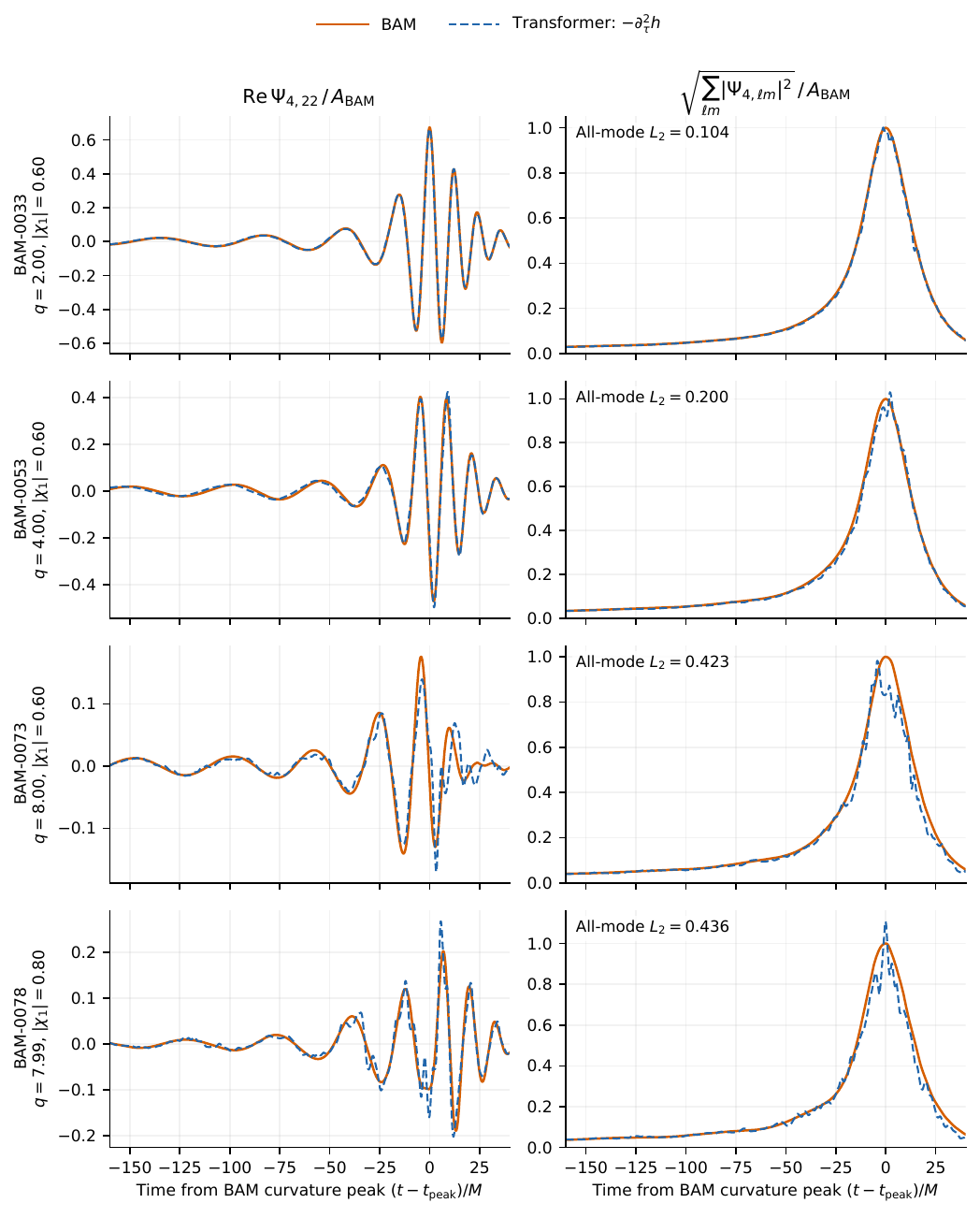}
\caption{Analytic model curvature $-\partial_\tau^2 h$ compared with native radius-$90M$ BAM $\Psi_4$ extractions for four precessing binaries. Left: real part of $\Psi_{4,22}$. Right: 21-mode amplitude norm, normalized by $A_{\rm BAM}=\max_t\sqrt{\sum_{\ell m}|\Psi^{\rm BAM}_{4,\ell m}(t)|^2}$, with the all-mode relative $L^2$ error on $[-160,40]M$ indicated.}
\label{fig:bam-curvature}
\end{figure*}

We also compare the analytic second derivative $-\partial_\tau^2h$ against six auxiliary precessing BAM simulations at extraction radius $R=90M$, with $q\simeq1,2,4,8$ and primary spin magnitudes near 0.6 or 0.8~\cite{Hamilton2024BAMCatalog}. All 21 native Weyl-scalar modes are aligned by a single time shift and rigid azimuthal rotation on $[-160,40]M$ (Fig.~\ref{fig:bam-curvature}). Relative curvature errors are $0.087$ and $0.104$ at $q\simeq1$ and 2, $0.200$ and $0.222$ for the two $q\simeq4$ cases, and $0.423$ and $0.436$ at $q\simeq8$, with 21-mode amplitude-envelope errors spanning $0.025$--$0.121$, even though the training objective contains no explicit $\Psi_4$ term.

%% file: tables/product_v6_paired.tex
\begin{tabular}{lrrrr}
\toprule
Comparator & Groups & $\Delta L_2(h)$ & $\Delta L_2(\Psi_4)$ & $\Delta\mathcal M$ (pp) \\
\midrule
\SEOB{} & 102 & $-0.0823\,[-0.1046, -0.0623]$ & $-0.1775\,[-0.2301, -0.1305]$ & $-1.3979\,[-2.0552, -0.8200]$ \\
\SurVTwo{} & 102 & $-0.0378\,[-0.0518, -0.0253]$ & $-0.0667\,[-0.1007, -0.0350]$ & $-0.4554\,[-0.7664, -0.2155]$ \\
\TPHM{} & 102 & $-0.1244\,[-0.1579, -0.0949]$ & $-0.2265\,[-0.2882, -0.1717]$ & $-2.4487\,[-3.6846, -1.3798]$ \\
\bottomrule
\end{tabular}

%% file: tables/product_v6_inbox.tex
\begin{tabular}{lrrrr}
\toprule
Model & $L_2(h)$ & Mean $\mathcal M$ (\%) & Median (\%) & P95 (\%) \\
\midrule
Transformer & 0.0184 & 0.0253 & 0.0163 & 0.0549 \\
\SEOB{} & 0.0520 & 0.2922 & 0.0740 & 1.8219 \\
\SurVTwo{} & 0.0240 & 0.0326 & 0.0082 & 0.0831 \\
\TPHM{} & 0.0746 & 0.3199 & 0.1280 & 1.1603 \\
\bottomrule
\end{tabular}

%% file: tables/product_v6_rit.tex
\begin{tabular}{lrrrr}
\toprule
RIT subset & Waveforms/groups & Mean $L_2(h)$ & Median & P95 \\
\midrule
All & 173/163 & 0.0809 & 0.0606 & 0.2892 \\
Aligned & 100/92 & 0.0525 & 0.0214 & 0.1577 \\
Precessing & 73/71 & 0.1176 & 0.0849 & 0.3271 \\
\bottomrule
\end{tabular}

%% file: tables/product_v6_reference.tex
\begin{tabular}{lrrrrr}
\toprule
Subset & Waveforms/groups & Fixed $L_2$ & With $R_z$ & Median $|\Delta\alpha|$ & Median $|\Delta t_{\rm peak}|/M$ \\
\midrule
All & 109/106 & 0.2414 & 0.1338 & $3.16^\circ$ & 0.66 \\
Aligned & 27/27 & 0.1952 & 0.0924 & $3.60^\circ$ & 0.61 \\
Precessing & 82/79 & 0.2572 & 0.1480 & $2.94^\circ$ & 0.67 \\
$q\leq4$ & 79/77 & 0.2286 & 0.1148 & $3.16^\circ$ & 0.61 \\
$q>4$ & 30/29 & 0.2756 & 0.1844 & $2.59^\circ$ & 0.82 \\
\bottomrule
\end{tabular}

%% file: sections/limitations.tex
\section{Regularity and physical scope}\label{sec:limitations}

\subsection{Parameter-space continuity}\label{sec:smooth}

\begin{figure*}[t]
\centering\includegraphics[width=\textwidth]{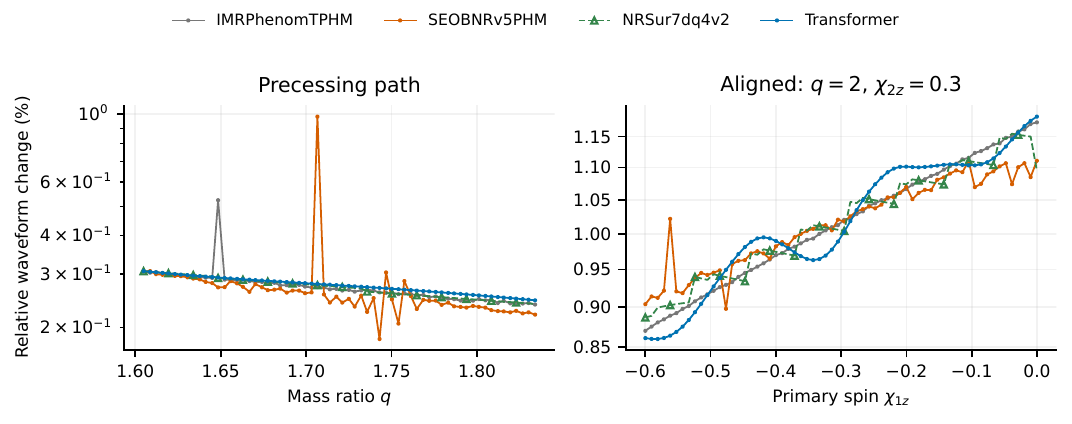}
\caption{Relative 21-mode strain changes for 64 uniformly spaced adjacent parameter pairs along a precessing path (left, step size $1/600$ of the segment) and an aligned-spin path (right, step size $\Delta\chi_{1z}=5\times 10^{-4}$), evaluated under a common strain-energy centering rule without pairwise time or phase alignment (Appendix~\ref{app:scans}). \SEOB{} includes its antisymmetric $(2,2)$, $(3,3)$, and $(4,4)$ modes, and \SurVTwo{} uses the reference-epoch calibration map of Eq.~\eqref{eq:sur_tref}.}
\label{fig:smoothness}
\end{figure*}

The analytic structure of the spline coefficient map and its $C^4$ time continuity are established in Appendix~\ref{app:regularity}. To examine numerical continuity across parameter space, Fig.~\ref{fig:smoothness} compares adjacent waveforms along a precessing and an aligned-spin segment under a shared strain-energy centering rule without pairwise time or phase alignment (Appendix~\ref{app:scans}). On both paths, the Transformer varies smoothly without generator kinks. At two points on the precessing segment, dyadic step-size refinements confirm linear scaling down to $\|\Delta x\|_2\sim10^{-7}$ for both the Transformer and \SurVTwo{} (Fig.~\ref{fig:bracket-refinement}).

\subsection{Physical scope}\label{sec:temporal}\label{sec:reference}\label{sec:coverage}

The spline domain $\tau\in[-1620,160]M$ covers the late inspiral, merger, and $100M$ post-peak ringdown evaluated on $[-1500,100]M$. Because physical starting frequency scales as $M^{-1}$ and increases with azimuthal index $|m|$, analyses of low-mass binaries that enter the detector band earlier than $-1500M$ can hybridize the predicted inertial modes onto an earlier post-Newtonian, \SEOB{}, or \TPHM{} inspiral over a matching early-inspiral window. In the late post-merger tail $(40,100]M$, where the signal decays by orders of magnitude, the mean SXS relative error is $\LateExtStrain$ (below \SEOB{} and \TPHM{}), though further suppressing small-amplitude late oscillations remains beneficial for precision ringdown studies. For comparison workflows extending past $100M$, \SurVTwo{} is bounded above by $t_{\rm sur}\leq100M$ ($\tau\leq90.5M$), whereas the Transformer evaluates natively across the full $100M$ ringdown.

In parameter space, precessing NR evaluation coverage is concentrated at $q\leq6$ (with 24 evaluation systems in $4<q\leq6$ and one in $6<q\leq8$). High mass ratios with large positive aligned spin remain sparsely sampled in existing catalogues and exhibit strong azimuthal sensitivity, as illustrated by the validation simulation SXS:BBH:2633 at $q=8$ ($\chi_{1z}=+0.58$, Appendix~\ref{app:observations}). While synthetic teachers extend pretraining coverage to $q\leq20$ and individual high-$q$ cases such as SXS:BBH:2629 ($q\simeq10$) and the BAM $q\simeq8$ curvature comparisons show encouraging agreement, expanding public precessing NR simulations at $q>6$ will enable systematic quantitative benchmarking in that regime.

%% file: sections/conclusion.tex
\section{Conclusion}\label{sec:conclusion}

We have presented a Transformer model that maps the seven intrinsic parameters of a precessing binary black hole directly to 512 quintic B-spline coefficients for all 21 signed spin-weighted spherical-harmonic strain modes through $\ell=4$ over $[-1620,160]M$. Pretraining jointly on NR simulations and synthetic \Sur{}, \SEOB{} (extending coverage to $q=20$), and perturbative teachers precedes SXS strain fine-tuning. The symmetry-projected spline representation guarantees exact binary-exchange and equatorial-reflection equivariance, $C^4$ time continuity, and analytic parameter and temporal derivatives.

On the full SXS evaluation sample of \ProductRows\ waveforms in \ProductGroups\ physical systems over $[-1500,100]M$, the mean balanced strain error is \ProductStrainMean\ and the mean detector mismatch at $120\,M_\odot$ is \ProductMismatchMean\%. Across common support, the model improves upon \SEOB{}, \SurVTwo{}, and \TPHM{} in mean and P95 error at all five tested total masses ($60$--$180\,M_\odot$), while within the \SurVTwo{} calibration domain ($q\leq4$, $\chi\leq0.8$) it achieves lower mean strain error ($\VSevenInboxStrain$ versus $\SurInboxStrain$) and comparable high-mass mismatch. The SXS strain-derived curvature error is \ProductCurvatureMean, native BAM curvature errors across $q\simeq1$--$8$ range from $0.087$ to $0.436$, and batched compiled \texttt{float64} evaluation over 3561 time samples reaches $\CompiledGPUThroughput\text{ waveforms/s}$ on an A100 GPU and $\CompiledCPUThroughput\text{ waveforms/s}$ on four CPU threads. Validating early-inspiral hybridization and refining unprofiled reference-frame alignment remain natural extensions for full-band parameter estimation.

%% file: sections/reproducibility.tex
\section*{Data and code availability}

NR waveforms are obtained from the public SXS, RIT, and Cardiff--BAM catalogues~\cite{Boyle2019SXS,Scheel2025SXS,Healy2022RIT,Bruegmann2008BAM,Hamilton2024BAMCatalog}. Synthetic training waveforms use \Sur{}~\cite{Varma2019NRSur7dq4}; evaluation comparisons use \SurVTwo{}~\cite{Ravishankar2026NRSur7dq4v2}, \SEOB{}~\cite{RamosBuades2023SEOBNRv5PHM,Estelles2025Asym} (\texttt{pyseobnr}~\cite{Mihaylov2023pyseobnr}), \TPHM{}~\cite{Estelles2022TPHM} (\texttt{LALSuite}~\cite{LALSuite,Wette2020SWIGLAL}), and \textsc{BHPTNRSur2dq1e3}~\cite{Rink2024BHPTNRSur2dq1e3} (\texttt{gwsurrogate}~\cite{Field2014surrogate}).

The waveform-generation and automatic-differentiation software is developed under the MIT license at \url{https://github.com/jdong745/merger-transformer}, together with the trained model weights ($2{,}564{,}276$ parameters), reference-epoch calibration maps, dataset partition ledger, and evaluation scripts.

\section*{AI use disclosure}
We used multiple models from the GPT and Claude families, as well as Gemini 4 Argon and Kimi K3.1 Pro, to assist with code development, analysis, and manuscript preparation. The author takes responsibility for the methods, results, and text.

\begin{acknowledgments}
We thank Benjamin D. Wandelt for hosting the computing hardware used in this work at the Johns Hopkins Whiting School of Engineering.
We thank Google Antigravity for providing complimentary access to Gemini 4 Argon.
We thank the Simulating eXtreme Spacetimes (SXS) Collaboration~\cite{Boyle2019SXS,Scheel2025SXS}, the Rochester Institute of Technology (RIT) numerical-relativity group~\cite{Healy2022RIT}, and the Cardiff University gravity group and BAM collaboration~\cite{Bruegmann2008BAM,Hamilton2024BAMCatalog} for making their numerical-relativity waveform catalogues publicly available. We also thank the developers and maintainers of \texttt{LALSuite}~\cite{LALSuite,Wette2020SWIGLAL}, \texttt{pyseobnr}~\cite{RamosBuades2023SEOBNRv5PHM,Estelles2025Asym,Mihaylov2023pyseobnr}, and \texttt{gwsurrogate}~\cite{Field2014surrogate,Varma2019NRSur7dq4,Ravishankar2026NRSur7dq4v2,Rink2024BHPTNRSur2dq1e3} for providing open-source waveform models and evaluation tools.
\end{acknowledgments}

%% file: sections/appendix_data.tex
\section{Numerical-relativity data preparation}\label{app:observations}

The dataset ledger tracks 3503 catalogue records with identifiers, physical-system groups, partition roles, quality flags, and upstream teacher membership, including 278 excluded records and 80 BAM curvature-only records.

\subsection{Teacher membership, auxiliary fits, and \SurVTwo{} domain comparison}\label{app:teacher-exposure}

We trace the upstream waveform-fitting sets, parent samples, and test-only uses of each teacher, propagating these identities through physical-system groups and identified reruns. Table~\ref{tab:teacher-exposure} summarizes membership across the NR partitions. Identified waveform-fitting and conservative parent-sample families are assigned to training or validation. External \Sur{} $q=6$ test cases that were not used in fitting are eligible for evaluation unless included in another teacher's fitting or parent set. All auxiliary regressions and frame-alignment models---the reference-epoch calibration polynomials $\widehat\Delta(x)$ and $\widehat d(x)$, and the time-independent $SO(3)$ alignment fit applied to precessing \SEOB{} training teachers---are fitted exclusively on training simulations, leaving zero evaluation groups in any auxiliary fit.

\input{tables/product_v6_teacher_exposure}

For \Sur{}, Ref.~\cite{Varma2019NRSur7dq4} specifies 1528 cases shared between waveform and remnant fits. Matching the official remnant-fit coordinates\footnote{The public file is \texttt{remnant\_fits/fit\_7dq4.h5} in \href{https://zenodo.org/records/3455886}{Zenodo record 3455886}.} to SXS metadata, including body exchange, identifies 1374 distinct SXS simulations and 154 exchanged copies. Group assignment propagates this membership to retained reruns and related group members. The external $q=6$ test simulations SXS:BBH:2164--2263 were used only for out-of-sample testing in Ref.~\cite{Varma2019NRSur7dq4} and are eligible for evaluation unless included in another teacher's fitting or parent set.

The \SurVTwo{} comparator has a published roster of 1378 SXS simulation IDs~\cite{Ravishankar2026NRSur7dq4v2}. Its four additional IDs relative to the original roster (SXS:BBH:0827, 0843, 0985, and 1488) are assigned to training, and none of the identified fitting families enter evaluation.

For aligned-spin \SEOB{} calibration, 415 of the 442 ancillary records in \texttt{NR\_simulations.json} from Ref.~\cite{Pompili2023SEOBNRv5HM} have identified public SXS counterparts; 27 identities remain unresolved. Propagation through our rerun-family grouping policy gives 558 training records in 372 groups and 30 validation records in 19 groups, with none in evaluation. The retained SXS:BBH:3629 rerun of SXS:BBH:0159 is assigned to training.

\textsc{BHPTNRSur2dq1e3} inherits NR calibration through 90 synthetic \textsc{NRHybSur3dq8} waveforms~\cite{Rink2024BHPTNRSur2dq1e3}. The latter's published fitting set contains 91 SXS simulations, SXS:BBH:1419--1509, augmented by body exchange~\cite{Varma2019NRHybSur3dq8}. Identified fitting families are restricted to the first two partitions; the separate long-waveform hybridization tests are recorded as test-only use.

The parent NR collection for the \SEOB{} antisymmetric-mode fits~\cite{Estelles2025Asym} comprises 1405 precessing second-catalogue simulations~\cite{Boyle2019SXS} and 118 additional simulations. Applying conservative parent-sample and identified-family restrictions assigns the 49 identified public parent-sample records to training (40 records, 36 groups) or validation (nine records, nine groups).

At high mass ratios, \texttt{pyseobnr} 0.3.6 precessing waveforms exhibit increasing radiation-axis misalignment relative to SXS NR. For the 6144 precessing \SEOB{} training teachers, an eight-parameter, time-independent active $SO(3)$ rotation model parameterized by $x$ was fitted exclusively on 43 SXS training simulations ($q\in[5,6]$), substantially reducing balanced strain errors against held-out SXS training simulations (by $\sim15\%$--$24\%$ across $q\in[5,9]$). This frozen rotation is applied to all 6144 precessing \SEOB{} native waveforms prior to spline interpolation, with the covariant $SO(3)$ rotation applied simultaneously to the rank-12 coprecessing observation operator. Both the 2000 aligned-spin \SEOB{} teachers and the \SEOB{} evaluation comparator use unmodified \texttt{pyseobnr} 0.3.6.

Within the validation set, SXS:BBH:2633 ($q\simeq8$, $\chi_{1z}=+0.58$, $|\boldsymbol\chi_{1\perp}|\simeq0.55$) shows a large numerical-resolution bifurcation: although its \texttt{Lev4} horizon metadata are self-consistent, \texttt{Lev3} and \texttt{Lev4} merge $108.6M$ ($87$ orbits) apart with in-plane spin azimuths separated by $\sim92^\circ$ (raw-frame balanced $L_2$ discrepancy $0.874$) in a corner lacking neighboring SXS training runs. SXS:BBH:2633 is therefore retained in the full 166-waveform validation record and excluded during validation checkpoint selection (165 selected validation waveforms in 149 groups).

Finally, because \SurVTwo{} terminates natively at $t_{\rm sur}=100M$ ($\tau=90.5M$ in our time coordinate) and requires zero-padding across $(90.5,160]M$ on $[-1500,100]M$, Table~\ref{tab:nrsur-domain-sensitivity} compares the full $[-1500,100]M$ metrics against an unpadded evaluation truncated to $[-1500,50]M$ (prediction grid $[-1620,90.5]M$). Model rankings and statistical differences remain consistent across both windows.

\input{tables/product_v6_exposure_sensitivity}

\subsection{Waveform preparation}

For SXS simulations, continuous interpolation of the apparent-horizon trajectories determines the reference orbital-frequency crossing $M\Omega=\pi Mf_{\rm ref}$. The horizon positions, their interpolation derivatives, and the spin vectors at that epoch define the orthonormal reference triad $(\boldsymbol n,\hat{\boldsymbol L}\times\boldsymbol n,\hat{\boldsymbol L})$, into which all spherical-harmonic modes are rotated by a single constant rotation.

RIT reference inputs are reconstructed from the released horizon snapshots. Reference spins and the directed orbital triad are transported to the nominal frequency $\omega_{\rm target}=\pi M_{\rm nom}f_{\rm ref}$ ($M_{\rm nom}=120\,M_\odot$, $f_{\rm ref}=20\,\mathrm{Hz}$) by integrating the quasicircular post-Newtonian/EOB spin-precession equations (independently of the waveform generation models evaluated in Sec.~\ref{sec:results}). The resulting reference parameters incorporate the waveform-clock and radiative-frame corrections, with spins, the directed triad, and all signed modes transformed together.

When initial orbital frequencies estimated from catalogue metadata yield a transport duration differing from the waveform merger time by $|\Delta T_{\rm w}|>15M$, or exhibit frame-orientation discrepancies relative to the radiation axis, we apply a bounded reference-epoch and triad alignment correction prior to pretraining (affecting 58 RIT pretraining waveforms in total).

For RIT strain preparation, aligned-spin waveforms use the published strain, whereas precessing waveforms use fixed-frequency double integration of the released $\Psi_4$ modes with cutoff $\omega_0=0.4(2\pi f_{22,\rm start})$, where $f_{22,\rm start}$ is the initial $(2,2)$-mode frequency, and a finite-radius extrapolation correction for modes whose published integration used a higher cutoff. In the extrapolation correction, each mode's effective published cutoff $\omega_{\rm pub}$ is estimated from the published strain and curvature; when $\omega_{\rm pub}>1.5\omega_0$, the curvature spectrum is multiplied by
\begin{equation}
 \frac{1-c_\ell/(Ri\omega_{\rm t})}
 {1-c_\ell/(Ri\omega_{\rm e})},\qquad
 c_\ell=\frac{(\ell-1)(\ell+2)}2,
\end{equation}
where $R$ is the extraction radius, $\omega_{\rm t}=\operatorname{sgn}(\omega)\max(|\omega|,\omega_0)$, and $\omega_{\rm e}=\operatorname{sgn}(\omega)\max(|\omega|,\omega_{\rm pub})$ (taking $\operatorname{sgn}(0)=+1$), before double integration at $\omega_0$. Both aligned and precessing RIT branches then subtract a Gaussian low-frequency component of width $0.4$ orbital cycles in orbital-phase time across all 21 modes.

For SXS strain, the training and evaluation target is
\begin{equation}
 H_{\rm target}=H-J_E[H;u_{\rm add}],
 \label{eq:sxs_memory}
\end{equation}
where $H$ is the catalogue strain and $J_E[H;u_{\rm add}]$ is the BMS energy-flux memory contribution computed from the supplied modes from the catalogue integration start time $u_{\rm add}$~\cite{Mitman2020memory}.

Three nonspinning RIT simulations at $q\simeq32$ cover only the merger interval $[-160,40]M$ and share one physical-system group. Quality cuts exclude incomplete merger waveforms, simulations with measured eccentricity $e>0.03$ or missing eccentricity metadata, and inconsistent repeat simulations. For example, RIT-1018 is excluded because its phase evolution disagrees with the other nonspinning $q\simeq32$ runs.

We also screen the SXS horizon trajectories for localized outward excursions in the binary separation and associated orbital-frequency pulses. Events such as SXS:BBH:0302 persist under changes in differentiation and sampling, but have no comparably strong feature in the rotation-invariant quadrupolar radiation frequency. We exclude these coordinate-pulse cases from reference-label construction. SXS:BBH:0058 is omitted because its coincident coordinate and radiative oscillations remain unresolved; SXS:BBH:0324 has repeated oscillations and lacks an eccentricity estimate.

For the six auxiliary BAM $\Psi_4$ comparisons~\cite{Hamilton2024BAMCatalog}, reference spins and the orbital triad are evaluated at the near-zone orbital-frequency crossing, with the orbital normal estimated from stored puncture velocities via a local cubic Savitzky--Golay filter of half-width $20M$. All 21 native radius-$90M$ $\Psi_4$ modes are rotated into this triad, expressed in relaxed-total-mass units with the tortoise delay subtracted, and centered at the peak of the total curvature amplitude.

%% file: tables/product_v6_teacher_exposure.tex
\begin{table*}[t]
\centering
\caption{Upstream calibration membership across the NR partitions (simulations/physical-system groups). Waveform-fitting and conservative parent/campaign categories use our physical-system group and rerun-family rules; categories overlap and are not additive. Test-only use is distinguished from fitting. The last three rows summarize the training-only reference-epoch calibration maps ($\widehat\Delta$ and $\widehat d$) and the $SO(3)$ alignment fit applied to the precessing \SEOB{} training teachers.}
\label{tab:teacher-exposure}
\begin{tabular}{lrrr}
\toprule
Calibration / reference set & Train & Val. & Eval. \\
\midrule
Identified waveform fitting (union) & 1831/1532 & 30/19 & 0/0 \\
\SEOB{} aligned fitting & 558/372 & 30/19 & 0/0 \\
Possible antisymmetric parent / rerun & 1430/1296 & 29/29 & 0/0 \\
Additional identified parent sample & 40/36 & 9/9 & 0/0 \\
Unresolved ancillary calibration family & 114/106 & 11/11 & 0/0 \\
\Sur{} external $q=6$ test & 62/61 & 19/17 & 21/20 \\
\textsc{NRHybSur3dq8} long-waveform test & 0/0 & 0/0 & 8/7 \\
\midrule
Input-clock calibration $\widehat\Delta$ & 798/783 & 0/0 & 0/0 \\
Teacher output clock $\widehat d$ & 1282/1220 & 0/0 & 0/0 \\
\SEOB{} precessing teacher $SO(3)$ fit & 43/43 & 0/0 & 0/0 \\
\bottomrule
\end{tabular}
\end{table*}

%% file: tables/product_v6_exposure_sensitivity.tex
\begin{table}[t]
\centering
\caption{Comparison between the full $[-1500,100]M$ evaluation window and the unpadded $[-1500,50]M$ interval lying completely within the native domain limit of \SurVTwo{} ($t_{\rm sur}\leq 100M$, i.e., $\tau\leq 90.5M$; prediction grid $[-1620,90.5]M$).}
\label{tab:product-v6-exposure-sensitivity}\label{tab:nrsur-domain-sensitivity}
\resizebox{\columnwidth}{!}{\begin{tabular}{lrrrr}
\toprule
Metric & \multicolumn{2}{c}{Transformer} & \multicolumn{2}{c}{\SurVTwo{}} \\
\cmidrule(lr){2-3}\cmidrule(lr){4-5}
 & $[-1500,100]M$ & $[-1500,50]M$ & $[-1500,100]M$ & $[-1500,50]M$ \\
\midrule
\multicolumn{5}{l}{Common support (105 waveforms / 102 groups)} \\
$L_2(h)$ & 0.0301 & 0.0301 & 0.0679 & 0.0678 \\
$\mathcal{M}(60\,M_\odot)$ (\%) & 0.0579 & 0.0564 & 0.4369 & 0.4172 \\
$\mathcal{M}(120\,M_\odot)$ (\%) & 0.1496 & 0.1490 & 0.6049 & 0.5817 \\
$\mathcal{M}(180\,M_\odot)$ (\%) & 0.2358 & 0.2352 & 0.6949 & 0.6680 \\
\midrule
\multicolumn{5}{l}{Calibration domain $q\leq4.01,\,\chi\leq0.8$ (69 waveforms / 68 groups)} \\
$L_2(h)$ & 0.0184 & 0.0184 & 0.0240 & 0.0240 \\
$\mathcal{M}(60\,M_\odot)$ (\%) & 0.0146 & 0.0133 & 0.0466 & 0.0470 \\
$\mathcal{M}(120\,M_\odot)$ (\%) & 0.0253 & 0.0251 & 0.0326 & 0.0333 \\
$\mathcal{M}(180\,M_\odot)$ (\%) & 0.0340 & 0.0338 & 0.0338 & 0.0334 \\
\bottomrule
\end{tabular}}
\end{table}

%% file: sections/appendix_seams.tex
\section{Regularity, automatic differentiation, and evaluation cost}\label{app:regularity}\label{app:runtime}

\begin{figure*}[tp]
\centering\includegraphics[width=\textwidth]{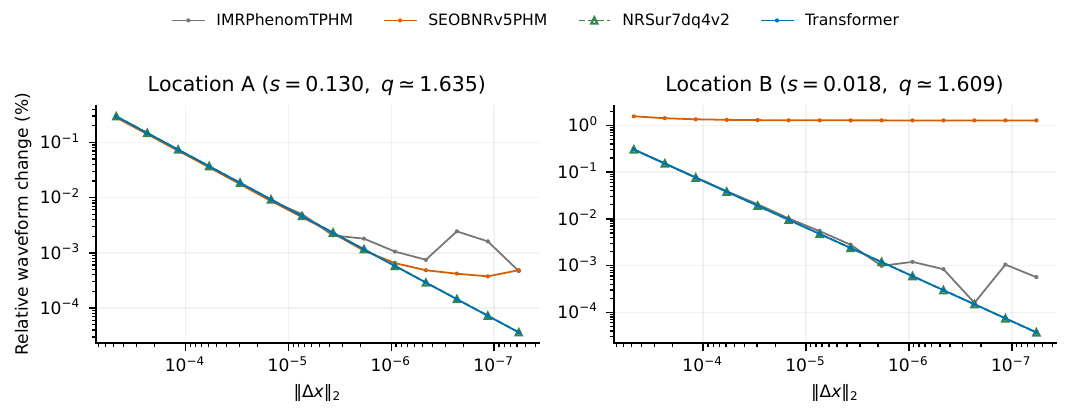}
\caption{Relative 21-mode strain change $\|H_{i+1}-H_i\|_2/\|H_i\|_2$ versus parameter step size $\|\Delta x\|_2$ over 14 dyadic step refinements at two fixed locations along the precessing path ($s=78/600\simeq0.130$, $q\simeq1.635$, and $s=11/600\simeq0.018$, $q\simeq1.609$).} \label{fig:bracket-refinement}
\end{figure*}

For fixed finite network weights, the input scaling~\eqref{eq:input_chart} and MLP/Transformer layers are compositions of real-analytic operations on $q>0$, and the symmetric-mass-ratio factor $s(\nu)$ in Eq.~\eqref{eq:nu-factor} is real analytic on $q>0$ and invariant under body exchange $q\mapsto q^{-1}$. Since the discrete group actions in the Reynolds average~\eqref{eq:symmetry_projection} are real linear, the resulting spline coefficients $c_{\ell m k}(x)$ are real analytic functions of $x=(q,\boldsymbol\chi_1,\boldsymbol\chi_2)$ for all $q>0$, including at $q=1$. In time, the degree-five B-spline basis with 506 simple interior bump knots has $C^4$ continuity for $h_{\ell m}(\tau;x)$ and $C^2$ continuity for $\Psi_{4,\ell m}(\tau;x)=-\partial_\tau^2 h_{\ell m}(\tau;x)$ on $(-1620,160)M$ (with one-sided derivatives at the clamped endpoints)~\cite{deBoor2001Splines}. Both the network layers and the spline basis are composed of elementary differentiable operations, so derivatives of these analytic expressions can be evaluated by automatic differentiation. Temporal derivatives at knots are subject to the continuity orders above.

Compiled \texttt{float64} forward evaluation agrees with uncompiled eager execution, satisfies discrete symmetry equivariance, and matches analytic spline differentiation to within $\CompiledMaxRel$ in relative norm across tested CPU and GPU architectures. For fast forward inference, the network and B-spline evaluation are compiled once via PyTorch/Inductor (\texttt{fullgraph=True}, \texttt{dynamic=False}, weight freezing enabled) in exact \texttt{float64}, while automatic differentiation remains supported through the eager interface. Each timed call evaluates the full parameter-to-strain pipeline, including four-way symmetry projection and 512-knot B-spline evaluation across all 21 signed modes at all 3561 time points on $[-1620,160]M$ after warm-up (excluding compilation and data loading). In batched steady-state evaluation, generating 1024 waveforms (microbatches of 128) on an NVIDIA A100-SXM4-40GB GPU takes $\CompiledGPUS\text{ s}$ ($\CompiledGPUAmortizedMS\text{ ms}$ per waveform, $\CompiledGPUThroughput\text{ waveforms/s}$, $\CompiledGPUMemory\text{ GiB}$ peak GPU memory), and generating a batch of \CompiledCPUBatch{} waveforms on \CompiledCPUThreads{} AMD EPYC 7773X CPU threads takes $\CompiledCPUBatchS\text{ s}$ ($\CompiledCPUAmortizedMS\text{ ms}$ per waveform, $\CompiledCPUThroughput\text{ waveforms/s}$).

\section{Moving-frame phase versus rigid inertial rotation}\label{app:registration}

Let $C(\tau)$ denote the coprecessing-frame modes and $R(\tau)\in SO(3)$ the coprecessing frame rotation. In the block-diagonal Wigner $D$-matrix representation $D[R(\tau)]$ for $\ell=2,3,4$, the inertial mode vector $h(\tau)=\{h_{\ell m}(\tau)\}$ is
\begin{equation}
 h(\tau)=D[R(\tau)]C(\tau).
\end{equation}
Applying a constant azimuthal phase shift $C_{\ell m}\mapsto e^{-im\alpha}C_{\ell m}$ in the coprecessing frame maps the inertial waveform to
\begin{equation}
 h(\tau)\mapsto D\!\left[R(\tau)R_z(\alpha)R(\tau)^{-1}\right]h(\tau).
\end{equation}
When the radiation axis precesses, $R(\tau)R_z(\alpha)R(\tau)^{-1}$ is a time-dependent rotation that alters the physical waveform. It cannot be absorbed into a single rotation of the inertial observer's reference frame. Our training and evaluation metrics therefore use the rigid inertial rotation in Eq.~\eqref{eq:alignment}.

Furthermore, an orbital phase extracted from the dominant coprecessing mode via $\phi=-\tfrac12\operatorname{unwrap}(\arg C_{22})$ is defined only modulo $\pi$: shifting $\phi\mapsto\phi+\pi$ flips the sign of every odd-$m$ coorbital mode $C_{\ell m}e^{im\phi}$. Predicting all 21 signed inertial modes directly avoids introducing this $\pi$-branch ambiguity during target construction.

\section{Parameter-scan conventions}\label{app:scans}

To examine parameter-space continuity without pairwise time or phase alignment, we evaluate neighboring waveforms along a precessing segment $x(s)=x_a+s(x_b-x_a)$ ($0\leq s\leq1$) connecting
{\small
\begin{align*}
x_a&\simeq(1.605,\,0.061,\,0.114,\,-0.031,\,0.015,\,0.203,\,0.568),\\
x_b&\simeq(1.834,\,0.093,\,0.100,\,-0.054,\,0.076,\,0.128,\,0.439),
\end{align*}}
and along an aligned-spin segment with $q=2$, $\chi_{2z}=0.3$, zero in-plane spins, and $\chi_{1z}\in[-0.6,0]$. Figure~\ref{fig:smoothness} samples 64 adjacent parameter pairs with step size $\Delta s=1/600$ along the precessing segment and $\Delta\chi_{1z}=5\times10^{-4}$ along the aligned-spin path; Fig.~\ref{fig:bracket-refinement} performs 14 successive dyadic step halvings at two fixed locations on the precessing path ($s=78/600\simeq0.130$, $q\simeq1.635$, and $s=11/600\simeq0.018$, $q\simeq1.609$, labeled Location A and Location B). The \SurVTwo{} curves use the reference-epoch calibration map of Eq.~\eqref{eq:sur_tref}. Each generated waveform is centered at its smooth strain-energy time moment
\begin{equation}
 t_E=\frac{\sum_j t_j w_j}{\sum_j w_j},\qquad
 w_j=\left(\frac{P_j}{\max_kP_k}\right)^8,
 \label{eq:scan_clock}
\end{equation}
where $P_j=\sum_{\ell m}|h_{\ell m}(t_j)|^2$. After interpolating $H(u)=h(u+t_E)$ onto a common time grid inside $[-180,20]M$ at $0.5M$ spacing, we compute the relative 21-mode difference $\|H_{i+1}-H_i\|_2/\|H_i\|_2$ (plotted as a percentage in Figs.~\ref{fig:smoothness} and~\ref{fig:bracket-refinement}).

Across the 14 dyadic refinements, the Transformer achieves linear scaling with fitted log--log slopes of $1.0000$ at both locations ($0.999986$ at A and $1.000002$ at B); at the finest step, $\|\Delta x\|_2\simeq5.77\times10^{-8}$, the relative waveform changes are $3.62\times10^{-7}$ and $3.72\times10^{-7}$, matching \SurVTwo{} ($3.63\times10^{-7}$ and $3.72\times10^{-7}$).

%% file: sections/appendix_method_checks.tex
\section{Derivative penalties and envelope differentiation}\label{app:gradient}

To qualify SXS waveforms for finite-difference supervision and curvature evaluation, we verify on the interior of the observed time interval that cubic and quintic spline second derivatives, as well as full and twofold-decimated quintic second derivatives, agree within $2\%$ relative $L^2$ error, and that cubic spline reconstruction of the native strain agrees with the $1M$-sampled strain within $10^{-5}$ relative $L^2$.

On the $0.5M$ strain grid ($\Delta\tau=0.5$), the full-window strain weights supplement the equal-weight core segments $W\in\{[-1000,-160),[-160,40]\}$ with the early inspiral segment $[-1500,-1000)$ (factor $0.1$, denominator floor $10^{-3}Q_{i,[-160,40]}$) and the late ringdown segment $(40,100]$ (target factor $0.1$, denominator floor $0.005\,Q_{i,[-160,40]}$, omitted for \textsc{BHPTNRSur2dq1e3}), blending smoothly from the merger sample density at $\tau=40M$ to $\tau\geq48M$ via the $C^2$ quintic smoothstep $S(u)=u^3(10-15u+6u^2)$ with $u=\operatorname{clip}_{[0,1]}[(\tau-40)/8]$.

For temporal regularization on the $0.5M$ grid, the finite-difference operators use the second-order central stencils in physical mass units,
\begin{align}
 (\Delta^{(1)}z)_j&=\frac{z_{j+2}-z_{j-2}}{4\,\Delta\tau},\\
 (\Delta^{(2)}z)_j&=\frac{z_{j+1}-2z_j+z_{j-1}}{(\Delta\tau)^2},
\end{align}
evaluated at stencil centers $\tau_j\in[-1499,99]M$ where all five points $j-2,\ldots,j+2$ are observed. For each segment $W$, let $J_W$ denote the valid stencil centers and $Q^{\rm fd}_{i,W}=\sum_{\ell m}\sum_{j\in J_W}|y^i_{\ell m,j}|^2$. The normalized $p$th-order finite-difference error on the core segments $W\in\{[-999,-160),[-160,39]\}$ is
\begin{equation}
 D_{i,W}^{(p)}=
 \frac{\displaystyle\sum_{\ell m}\sum_{j\in J_W}
 |\Delta^{(p)}(\mathcal A_{\delta,\alpha}h-y^i)_{\ell m,j}|^2}
 {\displaystyle\max\!\bigl[\sum_{\ell m}\sum_{j\in J_W}
 |\Delta^{(p)}y^i_{\ell m,j}|^2,10^{-12}Q^{\rm fd}_{i,W}\bigr]}.
\end{equation}
Early finite differences on $[-1499,-999)$ carry factor $0.1$ with normalizer floor $0.2$ times the early-core derivative denominator, while late finite differences on $(39,99]M$ carry target factor $0.1$, normalizer floor $0.02$ times the merger-core derivative denominator, and the smooth quintic onset $S(\operatorname{clip}_{[0,1]}[(\tau_j-39)/8])$. During joint pretraining ($0$--$32{,}000$), core and early $D^{(1)},D^{(2)}$ penalties receive unit scale while late $D^{(2)}$ carries scale $0.05$. During SXS-only fine-tuning, only the late-ringdown second-derivative penalty $D_i^{\rm late}$ on $(39,99]M$ is retained, with weight $\lambda_{\rm D2}=0.02$ through update $52{,}000$ and $\lambda_{\rm D2}=0.40$ through update $64{,}000$. For partial-mode synthetic teachers, the mode mask or covariant coprecessing subspace projection $\mathsf P_i(\tau_j)$ is applied prior to finite differencing.

With the merger-window strain energy on the $1M$ grid defined by
\begin{equation}
 Q_{i,\rm m}=\sum_{\ell m}\sum_{\tau\in[-160,40]}
 O_i(\tau)|y^i_{\ell m}(\tau)|^2,
\end{equation}
the curvature error on the $0.5M$ grid $W_{\rm cur}=[-160,40]M$ with valid-sample mask $O_i^{\Psi_4}$ is
\begin{equation}
 C_i=
 \frac{\displaystyle\tfrac12\sum_{\ell m}\sum_{\tau\in W_{\rm cur}}
 O_i^{\Psi_4}|\mathcal A_{\delta,\alpha}\Psi_{4,\ell m}-\psi^i_{\ell m}|^2}
 {\displaystyle\max\!\bigl[\tfrac12\sum_{\ell m}\sum_{\tau\in W_{\rm cur}}
 O_i^{\Psi_4}|\psi^i_{\ell m}|^2,10^{-12}Q_{i,\rm m}\bigr]},
\end{equation}
where the factor $1/2$ accounts for the $0.5M$ grid spacing relative to the $1M$ strain grid.

For the joint-stage ringdown tail loss on $W_{\rm tail}=(10,40]M$, let $p_i(\tau)=\sum_{\ell m}|y^i_{\ell m}(\tau)|^2$ and $p_{i,\max}=\max_{\tau\in[-160,40]}p_i(\tau)$. With $u=\operatorname{clip}_{[0,1]}[(\tau-10)/5]$ and $S(u)=u^3(10-15u+6u^2)$, the time weight
\begin{equation}
 w_i^{\rm tail}(\tau)=S(u)\operatorname{clip}_{[1,10]}
 \sqrt{\frac{p_{i,\max}}{\max[p_i(\tau),p_{i,\max}/100]}}
\end{equation}
ramps smoothly from $10M$ to $15M$ up to a cap of $10$, giving
\begin{equation}
 T_i=
 \frac{\displaystyle\sum_{\ell m}\sum_{\tau\in W_{\rm tail}}O_iw_i^{\rm tail}
 |\mathcal A_{\delta,\alpha}h_{\ell m}-y^i_{\ell m}|^2}
 {\displaystyle\max\!\bigl[\sum_{\ell m}\sum_{\tau\in W_{\rm tail}}O_iw_i^{\rm tail}
 |y^i_{\ell m}|^2,10^{-12}Q_{i,\rm m}\bigr]}.
\end{equation}
For each profiled objective---$H_i(\delta,\alpha)$ in Eq.~\eqref{eq:loss} or $H_i(\delta,\alpha)+0.1\gamma_iD_i(\delta,\alpha)+0.1T_i(\delta,\alpha)$ in Eq.~\eqref{eq:temporal_loss}---parameter gradients with respect to $\theta$ are evaluated at the minimizing alignment $(\delta^*,\alpha^*)$ via the envelope theorem~\cite{MilgromSegal2002Envelope}.

\section{Equivariance and azimuthal gauge}\label{app:gauge_sensitivity}

The real-linear mode transformations $D_g$ in Eq.~\eqref{eq:mode_symmetry} form a representation of $G=\{I,E,P,EP\}$ and the scalar factor $s(\nu)$ is $G$-invariant; reindexing the Reynolds average in Eq.~\eqref{eq:symmetry_projection} therefore gives $c(gx)=D_gc(x)$, and linearity over the real spline basis $B_k(\tau)$ carries this equivariance to $h_{\ell m}(\tau;x)$ and $\Psi_{4,\ell m}(\tau;x)$.

Discrete $G$-equivariance does not fix a continuous azimuthal gauge across generic parameter space: any loss minimized over a rigid $R_z(\alpha)$ rotation is invariant under a parameter-dependent azimuthal phase shift $h_{\ell m}(\tau;x)\mapsto e^{-im\alpha_0(x)}h_{\ell m}(\tau;x)$, which shifts the observer azimuth $\varphi$ in Eq.~\eqref{eq:mode_expansion}. The unaligned SXS anchor $A_i=H^{\rm core}_i(0,0)$ and the fixed-azimuth supervision on the 24{,}000 in-domain \Sur{} teachers constrain this continuous gauge during training. Table~\ref{tab:product-v6-reference} measures the remaining time and azimuth offsets.

%% file: sections/appendix_reference_inputs.tex
\par\vspace{0.65em}\kern0pt
\section{Reference-input conventions}\label{app:reference-inputs}

A nominal reference frequency $Mf_{\rm ref}$ can map to different physical times relative to merger across waveform generators whenever a model defines $f_{\rm ref}$ from waveform phase or post-Newtonian frequency rather than the NR apparent-horizon orbital frequency $\Omega$. For \Sur{}, we bridge this difference between the NR horizon reference epoch and the surrogate's internal frequency-to-time map $t_{\rm map}^{\rm Sur}(x)$ using two polynomial maps fitted on SXS training simulations (Appendix~\ref{app:teacher-exposure}).

First, on 798 SXS training simulations (783 physical groups) within the \Sur{} domain, we fit the difference between the NR horizon reference time (measured relative to the curvature-moment time origin $t_0$) and $t_{\rm map}^{\rm Sur}(x)$ using a 60-feature cubic ridge regression model $\widehat\Delta(x)$ (ridge penalty $0.01$). Evaluating \Sur{} at $t_{\rm ref}^{\rm Sur}(x)=t_{\rm map}^{\rm Sur}(x)+\widehat\Delta(x)$ while holding the horizon spin vectors $\boldsymbol\chi_{1,2}$ fixed aligns the surrogate's reference epoch with the SXS horizon convention. This map is used to generate the synthetic \Sur{} training sets over $[-1620,90.5]M$. For the \SurVTwo{} comparator in Sec.~\ref{sec:results}, we retain the resulting event-time prediction $\widehat T(x)=t_{\rm map}^{\rm Sur}(x)+\widehat\Delta(x)$ and initialize its dynamics at that time with the same spin vectors.

Second, to align the waveform time origin of the 24{,}000 in-domain \Sur{} teachers to the SXS target clock without profiling over $(\delta,\alpha)$ during training, we perform a time-only fit at fixed azimuth ($\alpha=0$) on 1282 SXS training simulations (1220 physical groups),
\begin{equation}
 h^{\rm SXS}_{\ell m}(\tau)\simeq
 h^{\rm Sur}_{\ell m}(\tau-d_i),\qquad \alpha=0,
\end{equation}
and fit $d_i$ with a 170-feature quartic ridge polynomial $\widehat d(x)$ ($\alpha_{\rm ridge}=0.1$, five-fold group cross-validation RMSE $1.093M$, training RMSE $0.931M$). Both $\widehat\Delta(x)$ and $\widehat d(x)$ use standardized Cartesian polynomial features of even total degree in the in-plane spin components and an unpenalized intercept. During joint pretraining, the 24{,}000 in-domain \Sur{} teachers are evaluated on $[-1500,100]M$ at $\alpha=0$ and $\delta=-\widehat d(x)$ in Eq.~\eqref{eq:alignment} (with $t_{\rm sur}>100M$ masked out), whereas all other synthetic teachers optimize $(\delta,\alpha)$ freely.